\documentclass[10pt,journal,epsfig,twoside]{IEEEtran}
\usepackage{graphicx}
\usepackage{subfigure,color}
\usepackage{amssymb}
\usepackage{cite,comment}
\usepackage{amsmath}
\usepackage{hhline}
\usepackage{bm,comment}
\usepackage{algorithm}
\usepackage{fancyhdr,balance}
\usepackage{algpseudocode}
\usepackage{multirow}
\makeatletter

\newcommand{\Rmnum}[1]{\expandafter\@slowromancap\romannumeral #1@}
\newcommand{\mv}[1]{\mbox{\boldmath{$ #1 $}}}

\makeatother

\begin{document}
\title{Intelligent Reflecting Surface Aided Wireless Networks: From Single-Reflection to Multi-Reflection Design and Optimization}
\author{Weidong Mei,~\IEEEmembership{Member,~IEEE}, Beixiong Zheng,~\IEEEmembership{Member,~IEEE}, Changsheng You,~\IEEEmembership{Member,~IEEE}, and Rui Zhang,~\IEEEmembership{Fellow,~IEEE}
\thanks{W. Mei and R. Zhang are with the Department of Electrical and Computer Engineering, National University of Singapore, Singapore 117583 (e-mail: \{wmei, elezhang\}@nus.edu.sg). {\it (Corresponding author: Rui Zhang.)}}
\thanks{B. Zheng is with the School of Microelectronics, South China University of Technology, Guangzhou 511442, China. He was with the Department of Electrical and Computer Engineering, National University of Singapore, Singapore 117583 (e-mail: bxzheng@scut.edu.cn).}
\thanks{C. You is with the Department of Electronic and Electrical Engineering, Southern University of Science and Technology (SUSTech), Shenzhen 518055, China. He was with the Department of Electrical and Computer Engineering, National University of Singapore, Singapore 117583 (e-mail: youcs@sustech.edu.cn).}}
\IEEEspecialpapernotice{(Invited Paper)}
\markboth{\textsc{Proceedings of the IEEE}}{\textsc{Proceedings of the IEEE}}
\maketitle

\begin{abstract}
Intelligent reflecting surface (IRS) has emerged as a promising technique for wireless communication networks. By dynamically tuning the reflection amplitudes/phase shifts of a large number of passive elements, IRS enables flexible wireless channel control and configuration, and thereby enhances the wireless signal transmission rate and reliability significantly. Despite the vast literature on designing and optimizing assorted IRS-aided wireless systems, prior works have mainly focused on enhancing wireless links with {\it\textbf{single signal reflection}} only by one or multiple IRSs, which may be insufficient to boost the wireless link capacity under some harsh propagation conditions (e.g., indoor environment with dense blockages/obstructions). This issue can be tackled by employing two or more IRSs to assist each wireless link and jointly exploiting their single as well as {\it\textbf{multiple signal reflections}} over them. However, the resultant double-/multi-IRS aided wireless systems face more complex design issues as well as new practical challenges for implementation as compared to the conventional single-IRS counterpart, in terms of IRS reflection optimization, channel acquisition, as well as IRS deployment and association/selection. As such, a new paradigm for designing {\it\textbf{multi-IRS cooperative passive beamforming}} and {\it\textbf{joint active/passive beam routing}} arises which calls for innovative design approaches and optimization methods. In this paper, we give a tutorial overview of multi-IRS aided wireless networks, with an emphasis on addressing the new challenges due to multi-IRS signal reflection and routing. Moreover, we point out important directions worthy of research and investigation in the future.
\end{abstract}
\begin{IEEEkeywords}
Intelligent reflecting surface (IRS), smart radio environment, double-IRS system, multi-IRS system, passive reflection optimization, cooperative passive beamforming, joint active/passive beam routing, channel estimation.	
\end{IEEEkeywords}

\section{Introduction}
While the fifth-generation (5G) wireless network is being actively deployed worldwide, both academia and industry have started envisaging and planning  the next-generation (6G) wireless network to more capably support the fast-growing demands for emerging applications, such as pervasive virtual/augmented reality, autonomous driving, tactile Internet, robotic communications, etc\cite{saad2019vision}. These applications impose new and more stringent wireless communication requirements than 5G in all of the key metrics including data rate, coverage, connectivity,  reliability, and latency. Moreover, the energy and spectral efficiency of 6G network is expected to be further improved to attain a sustainable network  capacity growth in the future cost-effectively. Furthermore, new techniques need to be developed to endow future wireless systems with new functionalities such as high-precision radio frequency (RF) sensing \cite{zong20196g}, coexistent wireless information and power transmissions \cite{clerckx2019fund}, as well as terrestrial and aerial communications \cite{zeng2019accessing}. The above new  requirements and functions of 6G, however, may not be fully achieved by current  5G techniques \cite{boccardi2014five}, e.g., ultra-dense network, millimeter wave (mmWave) communications, and massive multiple-input multiple-output (MIMO), which generally will incur increasingly higher energy and hardware costs as well as more challenging interference and backhaul/fronthaul issues in 6G.

\begin{figure*}[t]
\begin{center}
\includegraphics[width=16.5cm]{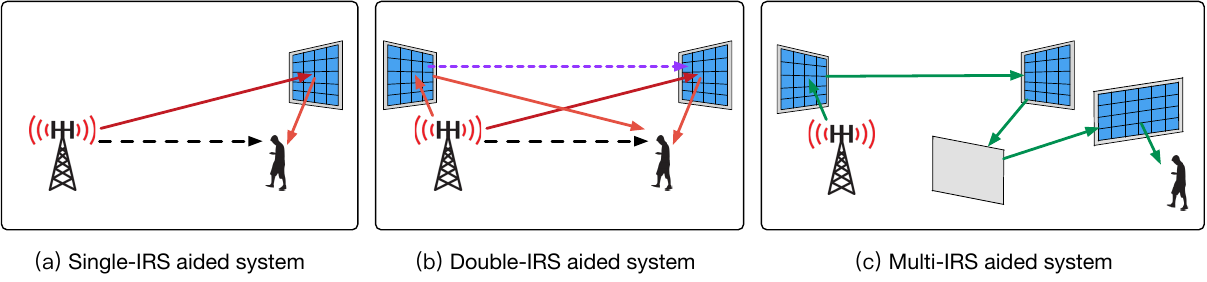}
\caption{Wireless communication systems aided by: (a) a single IRS with the direct and one single-reflection links; (b) two/double IRSs with the direct, two single-reflection, and one double-reflection links; and (c) multiple IRSs with their successive  multi-reflection links.}
\label{Fig:Three_sys}
\end{center}
\vspace{-12pt}
\end{figure*}

Recently, \emph{intelligent reflecting surface} (IRS) \cite{JR:wu2018IRS,JR:wu2019IRSmaga,wu2021intelligent} (or its various equivalents such as reconfigurable intelligent surface (RIS) \cite{basar19_survey,di2020smart_JSAC}) has emerged as a promising technique to proactively control the radio propagation environment by smart signal reflection and thereby enhance the wireless communication performance significantly. Specifically, IRS is a digitally controlled metasurface that comprises a large number of passive reflecting elements, each being  capable of changing the amplitude and/or phase of the incident signal independently \cite{liaskos2018new}. With IRSs properly deployed in wireless networks, the wireless channels can thus be dynamically adjusted by jointly tuning the reflecting elements of all IRSs so as to maximize the communication throughput. Note that this approach is in sharp contrast to traditional wireless techniques that can only compensate for or adapt to the wireless channel fading, while the wireless channels still remain largely random and uncontrolled. For example, IRS can be utilized to bypass obstacles/obstructions in wireless channels \cite{JR:wu2018IRS}, refine their realizations/distributions \cite{huang2021transforming}, improve the multi-antenna/multi-user channel rank condition \cite{ozdogan2020using}, etc. 
 Furthermore, since IRS operates in full-duplex mode with passive reflection only,  it is exempt from  signal amplification/processing noise and hence greatly enhances the spectral efficiency as compared to active relays \cite{ntontin2019reconfigurable}. Last but not least, from an implementation viewpoint, IRS is of light profile and weight as well as low energy consumption, thus it can be easily coated on environmental objects with low-cost energy supplies (e.g., battery). 

The appealing advantages of IRS have spurred intense  research interests in recent years. Existing  works on IRS design and reflection optimization  have mainly focused on the basic setup of single-IRS aided wireless systems as shown in Fig.~\ref{Fig:Three_sys}(a), where an IRS is usually deployed at the side of users to enhance the local  communication coverage and rate performance. Besides, an alternative single-IRS deployment strategy places  the IRS near the base station (BS) or access point (AP), such that the BS/AP with only a moderate number of antennas can achieve comparable communication performance to a massive MIMO BS/AP, by utilizing the nearby IRSs to enable fine-grained passive beamforming towards its served users \cite{zhang2021intelligent}. Note that these two single-IRS deployment strategies are consistent with the theoretical finding  that the IRS should be deployed near either the transmitter or the receiver to minimize the  severe product-distance path-loss over the two links with its assisted  BS and user \cite{wu2021intelligent}. Nevertheless, employing  one single IRS for each wireless link in general has limited control over its wireless channel  and thus may not be able to unlock the full potential of IRS in communication performance enhancement due to the following reasons. First, each IRS can serve users effectively only if they reside in its reflection half-space, which results in  limited coverage of one single IRS \cite{you2022deploy}. Second, even when the user is located in one IRS's reflection half-space, a blockage-free link between the BS and user via the IRS may not be always available in practice, since the single-reflection path provided by one single IRS may fail to bypass obstacles in complex environment, e.g., multi-turn corridors in an indoor environment. Third, due to practical constraints on the maximum size of each IRS,  the single IRS may only offer limited passive beamforming gains for each user, thus cannot help boost its achievable transmission rate drastically. Last, as the channels between different reflecting elements of each IRS and its assisted BS/users are usually correlated, the single-IRS aided system may suffer a limited spatial multiplexing gain due to the resultant  low-rank multi-antenna/multi-user channels.

To overcome the aforementioned limitations of single-IRS aided communications, new research efforts have been recently devoted to designing more efficient \emph{multi-IRS} aided wireless systems with two or more IRSs engaged in assisting each wireless link. Specifically, for the \emph{double-IRS} aided communication system shown in Fig.~\ref{Fig:Three_sys}(b), two IRSs can be deployed at the sides of BS and  users, respectively, to enhance their communitarian performance \cite{you2022deploy,han2020cooperative,zheng2021double,zheng2021uplink,zheng2021efficient,you2021wireless,dong2021double,kang2021irs,han2021double,shao2021joint,abdullah2021double,kim2021multi}. Note that besides the direct BS-user link and two single-reflection links each provided by one of the two IRSs, the double-IRS aided system provides a new \emph{double-reflection} link across the two IRSs for serving each user (see Fig.~\ref{Fig:Three_sys}(b)), which endows more degrees-of-freedom (DoFs) for refining the overall wireless channel, especially when the direct and two single-reflection links are all blocked \cite{you2021wireless}. Particularly, it has been shown in \cite{han2020cooperative} that if the double-reflection channel is  line-of-sight (LoS), it can achieve a higher scaling order of the passive beamforming gain than each single-reflection link, as the number of total reflecting elements, $N$, becomes asymptotically large, i.e., $\mathcal{O}(N^4)$ versus $\mathcal{O}(N^2)$.
This is fundamentally due to the more dominant cooperative passive beamforming (CPB) gain over the inter-IRS LoS channel. Motivated by the promising performance gains of double-IRS system, more than two IRSs can be employed to further enhance the performance of each wireless link by deploying more IRSs in the wireless network and properly assigning them to help different links at the same time \cite{mei2021performance,li2020weighted,zhang2019analysis,yang2021energy,he2021cooperative,mei2021cooperative,mei2021massive,mei2021mbmh,mei2021distributed,huang2021multi,zhang2021weighted,nguyen2021managing}. For example, as shown in Fig.~\ref{Fig:Three_sys}(c), each user may communicate with its serving  BS over a \emph{multi-reflection} links where the signal is successively reflected by a set of selected IRSs \cite{mei2021cooperative,mei2021massive,mei2021mbmh,mei2021distributed,huang2021multi,zhang2021weighted,nguyen2021managing}. Compared to the single- and double-reflection links, the multi-reflection link in general provides more DoFs to bypass dense and scattered obstacles in complex environment, via proper IRS selection/association. Moreover, the multi-reflection link offers an even higher order of CPB gain than the double-reflection link, which can counteract the product-distance path-loss that also increases with the number of IRS reflections \cite{mei2021cooperative}. Furthermore, the diverse reflection paths available in the multi-IRS aided system provides more path diversity that helps achieve higher spatial multiplexing gains than both the single- and double-IRS aided systems for supporting multiple users, as well as endows more flexibility in selecting the reflection paths for different users to meet their respective quality-of-service (QoS) requirements. In summary, we compare the key performance metrics of  the single-, double-, and multi-IRS reflection links in Table~\ref{Table:Comp}.
\begin{table*}[t]
\centering
{\caption{Comparison of different types of IRS reflection links}\label{Table:Comp}
\normalsize\resizebox{0.85\textwidth}{!}{
\begin{tabular}{|c|c|c|c|c|c|}
\hline
&\multirow{1}{*}{\textbf{Single-reflection link}}&\multirow{1}{*}{\textbf{Double-reflection link}}&\multirow{1}{*}{\textbf{Multi-reflection link }}\\
\hline
\textbf{Ability to bypass blockage} & Good & Very good & Excellent  \\
\hline
\textbf{Path diversity} & Moderate  &  High & Very high \\
\hline
\textbf{Passive beamforming gain} & Moderate  &  High & Very high  \\
\hline
\textbf{End-to-end path-loss} & Moderate  &  High & Very high \\
\hline
\end{tabular}}
}
\end{table*}

Although the double- and multi-IRS aided systems provide appealing advantages over their single-IRS counterpart, they are more complex and thus face new and more challenging issues in communication design and performance optimization, which are discussed as follows. 
 \begin{itemize}
 \item First, the IRS passive reflection design for the double- and multi-IRS aided systems is much more involved than that of the single-IRS counterpart. It is worth noting that for the single-IRS case with only one single-reflection link, the IRS passive reflection only needs to be jointly designed with the BS's active beamforming for performance optimization  (see, e.g., \cite{JR:wu2018IRS,huang2018largeRIS,zhang2020capacity,yu2020robust,MIMO_Pan,zhou2020spectral,yan2020passive}). However, for double-IRS aided system, it is necessary to jointly design the CPB of both the two IRSs together with the BS's active beamforming, due to  the presence of not only two single- but also one additional double-reflection links \cite{zheng2021double}. Moreover, the joint active/passive beamforming design becomes more challenging in the multi-user setup, since it is intricately coupled with the design of IRS-user association where each user may be associated with no IRS, one single IRS, or both IRSs \cite{you2022deploy}. As for the multi-IRS aided system, besides more complicated CPB design over more than two IRSs, a new challenge arises due to the selection of IRSs for each user. Specifically, it is essential to design an efficient \emph{IRS beam routing} solution for each user to select a set of IRSs between it and its serving BS for maximizing its end-to-end link capacity with the BS\cite{mei2021cooperative}. To this end, a key issue hinges on optimally balancing the fundamental  trade-off between maximizing the CPB gain and minimizing the end-to-end path-loss in the multi-IRS  beam routing design. Furthermore, in multi-user systems, the IRS beam routing design needs to cater to different QoSs of the users as well as their locations, which thus calls for efficient solutions  to assign the IRSs over different users to maximize their overall throughput, while mitigating the potential inter-user interference in the multi-user beam routing design.
\item Second, to maximize the performance gains in IRS-aided communications, it is indispensable but more practically  challenging to acquire the channel state information (CSI) in the double- and multi-IRS aided systems than the single-IRS case. Specifically, based on the existing works on single-IRS channel estimation, the two separate channels between the IRS and BS/user can be individually estimated for the semi-passive IRS equipped with additional sensing devices (see, e.g., \cite{taha2021enabling,alexandropoulos2020hardware,taha2020deep}), or their cascaded channel can be estimated for the fully-passive IRS without any sensors installed (see, e.g., \cite{OFDM_Protocol,OFDM_BX,jensen2019optimal,he2019cascaded,you2019progressive,zheng2020intelligent,zheng2020fast,you2020fast,liu2019matrix,wang2019channel,he2020channel,chen2019channel,wan2020broadband}). However, these methods are generally inapplicable to the double-IRS aided system due the co-existence of both the single- and double-reflection links as shown in Fig.~\ref{Fig:Three_sys}(b), which are intricately coupled and also entail more channel coefficients for estimation \cite{zheng2021efficient}. Furthermore, for multi-IRS aided systems, the channel estimation issue becomes even more challenging as multiple IRSs are present which results in more inter-IRS channels to be estimated \cite{mei2021distributed}. Thus, new and efficient channel estimation/acquisition methods for double-/multi-IRS systems need to be devised in order to achieve their practical CPB gains with low channel training overhead.  
\end{itemize}

Motivated by the above, this paper provides a tutorial overview of multi-IRS aided wireless networks, with an emphasis on addressing the new and more challenging issues in the IRS reflection optimization and channel acquisition design. Note that this paper  significantly differs from the existing survey/tutorial articles on IRS that mainly considered the IRS-aided wireless systems with single-reflection links only  (see, e.g., \cite{JR:wu2019IRSmaga,wu2021intelligent,basar19_survey,di2020smart_JSAC,elmossallamy2020reconfigurable,liu2021reconfigurable,huang2019holographic,gong2020toward,bjornson2020reconfigurable,bjornson2021reconfigurable,you2021enabling,zheng2021survey}). Specifically, this paper presents a general system model for the  multi-IRS aided wireless network, and pursues  a systematic design framework to address two key issues in the double- and multi-IRS aided communication systems, namely, IRS passive reflection optimization and channel acquisition. We first consider the double-IRS system to draw essential insights into the optimal design given two IRSs, and then address the more challenging multi-IRS system with IRS selection and beam-routing optimization. Moreover, we discuss open issues in the design of double-/multi-IRS aided wireless network to inspire future research in this important new direction of IRS.   

The rest of this paper is organized as follows. Section \Rmnum{2} presents the general  system model for a multi-IRS aided wireless network, including its signal and channel models, as well as practical constraints. In Section \Rmnum{3}, we focus on the double-IRS aided system and address its design issues in passive reflection optimization and channel acquisition. These issues are further investigated for multi-IRS aided systems in Section \Rmnum{4}. Finally, we conclude this paper in Section \Rmnum{5}.

The following notations are used in this paper. Bold symbols in capital letter and small letter denote matrices and vectors, respectively. The conjugate, transpose, conjugate transpose, and rank of a vector or matrix are denoted as ${(\cdot)}^{*}$, ${(\cdot)}^{T}$ and ${(\cdot)}^{H}$, and ${\rm rank}(\cdot)$, respectively. ${\mathbb{R}}^n$ (${\mathbb{C}}^n$) denotes the set of real (complex) vectors of length $n$. For a complex number $s$, $s^*$, $\lvert s \rvert$ and $\angle s$ denote its conjugate, amplitude and phase, respectively. For a vector ${\mv a} \in {\mathbb{C}}^n$, ${\rm diag}({\mv a})$ denotes an $n \times n$ diagonal matrix whose entries are given by the elements of $\mv a$; while for a square matrix ${\mv A} \in {\mathbb{C}}^{n \times n}$, ${\rm diag}({\mv A})$ denotes an $n \times 1$ vector that contains the $n$ diagonal elements of ${\mv A}$. $\lVert \mv a \rVert$ denotes the Euclidean norm of the vector $\mv a$. $\lceil \cdot \rceil$ denotes the smallest integer larger than or equal to its argument. $\lvert A \rvert$ denotes the cardinality of a set $A$. $j$ denotes the imaginary unit, i.e., $j^2=-1$. For two sets $A$ and $B$, $A \cup B$ denotes the union of $A$ and $B$, and $A \backslash B$ denotes the set of elements that belong to $A$ but are not in $B$. $\emptyset$ denotes an empty set. ${\cal O}(\cdot)$ denotes the order of convergence or complexity. 

\section{System Model}\label{SysModel}
\begin{figure}[!t]
\centering
\includegraphics[width=3.2in]{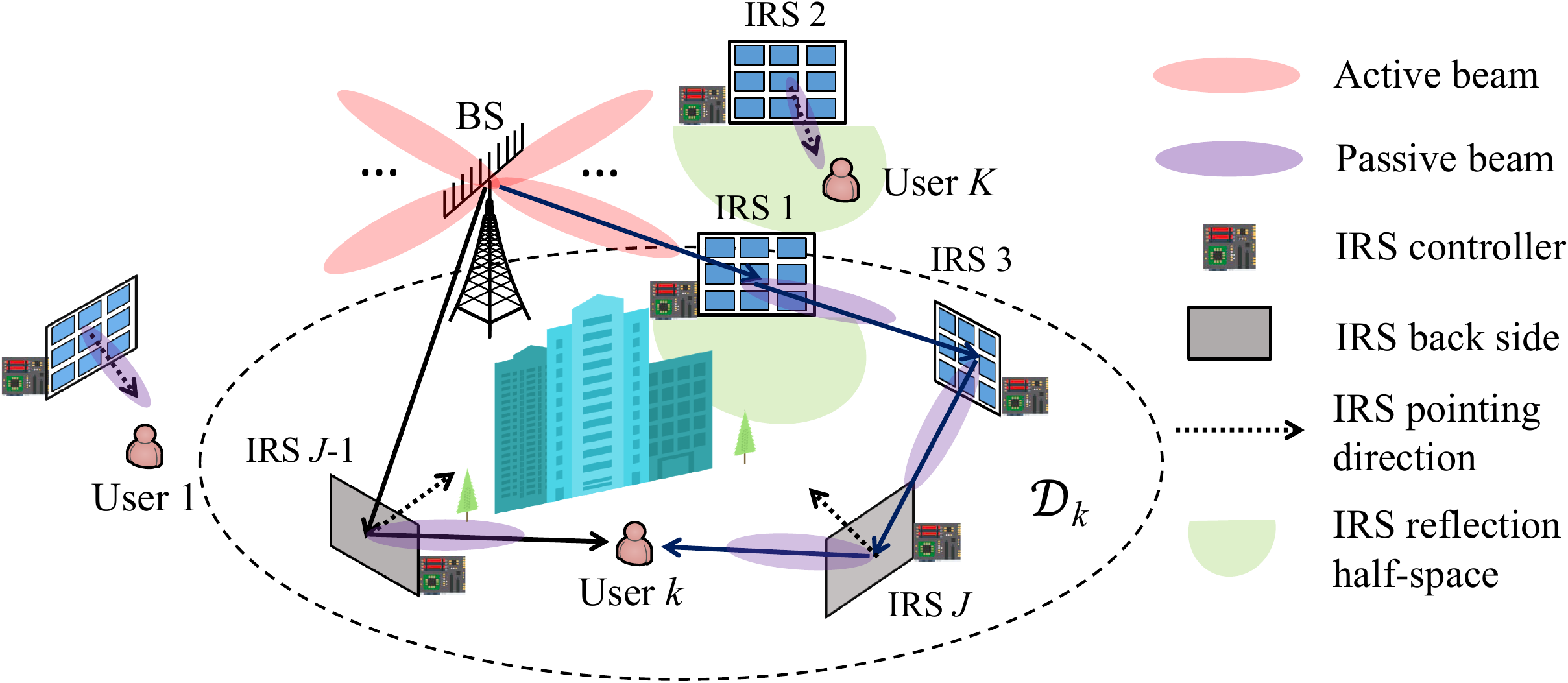}
\DeclareGraphicsExtensions.
\caption{A multi-IRS aided wireless network with joint BS/IRS active/passive beamforming.}\label{RefRegion}
\vspace{-9pt}
\end{figure}
Fig.\,\ref{RefRegion} depicts the downlink communications in a single-cell multi-user wireless system, where $J$ distributed IRSs are deployed to assist in the communications between a multi-antenna BS and $K$ single-antenna users. Note that the system model can also be applied to the uplink communications as well as extended to apply for more general setups with multiple cells and/or multi-antenna users. Assume that the BS is equipped with $N_B \ge K$ active antennas, while each IRS is equipped with $M$ passive reflecting elements. Moreover, a smart controller is attached to each IRS to dynamically tune its reflection coefficients as well as exchanging useful information (on e.g., CSI and control message) with the BS/users and other IRS controllers via separate and reliable wireless links. Without loss of generality, we focus on one snapshot (time slot) of the network with all the channels involved being given and fixed, while the results can be similarly applied over different time slots. For the ease of exposition, we consider that all channels are frequency-flat, while leaving the extension to the more general frequency-selective channels for our future work.\footnote{It is worth noting that IRS reflection and channel acquisition have been thoroughly studied for single-IRS systems (see, e.g., \cite{taha2021enabling,alexandropoulos2020hardware,taha2020deep,OFDM_Protocol,OFDM_BX,jensen2019optimal,he2019cascaded,you2019progressive,zheng2020intelligent,zheng2020fast,you2020fast,liu2019matrix,wang2019channel,he2020channel,chen2019channel,wan2020broadband}), while their extensions to the more complex double-/multi-IRS systems remain largely unexplored yet.} For convenience, we denote the sets of users and IRSs (or IRS controllers) as ${\cal K}\triangleq \{1,2,\cdots,K\}$ and ${\cal J}\triangleq \{1,2,\cdots,J\}$, respectively. Furthermore, the BS and user $k, k \in \cal K$ are labelled as nodes 0 and $J+k$ in the system, respectively. Note that each IRS can achieve 180$^\circ$ half-space reflection only in practice, i.e., only the signal impinging on its front half-space can be reflected. As such, we define a pointing direction for each IRS which is perpendicular to the IRS surface plane and points into its reflection half-space, as shown in Fig.\,\ref{RefRegion}.

Let $d_{i,j}, i \ne j$ denote the nominal distance between two nodes $i$ and $j$, for which some reference points can be selected at the BS/IRS for convenience. For each user $k, k \in \cal K$, we assume that only the IRSs in a region ${\cal D}_k$ may have an effect on its channel with the BS, as shown in Fig.\,\ref{RefRegion}; while all the other IRSs outside ${\cal D}_k$ are treated as random scatterers in the environment. Let $J_k \,(J_k \le J)$ denote the number of IRSs in ${\cal D}_k$. Accordingly, we define ${\mv f}_k  \in {\mathbb C}^{N_B \times 1}, k \in \cal K$ as the direct channel between the BS and user $k$, which accounts for all environment scatterers including the IRSs outside ${\cal D}_k$ and is free of signal reflections by the $J_k$ IRSs in ${\cal D}_k$. In addition, we define ${\mv Q}_{0,j} \in {\mathbb C}^{M \times N_B}, j \in {\cal J}$ as the channel from the BS to IRS $j$, ${\mv g}_{j,J+k}^{H} \in {\mathbb C}^{1 \times M}, j \in {\cal J}$ as that from IRS $j$ to user $k$, and ${\mv S}_{i,j} \in {\mathbb C}^{M \times M}, i,j \in {\cal J}, i \ne j$ as that from IRS $i$ to IRS $j$. To maximize the reflected signal power by each IRS, the reflection amplitude of all its elements is set to the maximum value of one. Hence, the reflection coefficient matrix and passive beamforming vector of each IRS $j, j \in \cal J$ is given by ${\mv \Phi}_j={\rm diag}\{e^{j\theta_{j,1}},\cdots,e^{j\theta_{j,M}}\} \in {\mathbb C}^{M \times M}$ and ${\mv \theta}_j = {\rm diag}({\mv \Phi}_j) \in {\mathbb C}^{M \times 1}$, respectively, with ${\theta_{j,m}}$ denoting the phase shift by the $m$-th element of IRS $j$, $m=1,2,\cdots,M$, $j \in \cal J$. 

In the presence of multiple IRSs in ${\cal D}_k$, the BS can communicate with each user $k$ through both their direct link (i.e., ${\mv f}_k$) and the reflected links by the IRSs in ${\cal D}_k$ with different number of reflections. For each reflected link, as an IRS can only achieve 180$^\circ$ signal reflection, the BS and user $k$ should be located in the reflection half-space of their respective next and previous reflecting IRSs, so as to achieve effective signal reflection. Furthermore, if the reflected link constitutes multiple IRSs, each intermediate IRS should be located in the reflection half-spaces of both its previous and next reflecting IRSs. For example, in Fig.\,\ref{RefRegion}, IRS 1 and IRS 2 cannot successively reflect the signal from the BS as they do not meet the above condition. Moreover, due to the substantial multiplicative path loss of the multi-reflection link, we only consider the paths where the BS’s signal is reflected from one IRS to a farther IRS from the BS. For example, in Fig.\,\ref{RefRegion}, among all reflection paths from the BS to user $k$, we do not consider the path going through IRSs 3, 1 and $J$ successively, as it results in much higher cascaded path loss than that going through IRSs 1, 3, and $J$ successively. Under the above conditions, denote by $\Lambda_{k,n}$ the set of all $n$-reflection ($n \ge 1$) paths from the BS to user $k$ via the IRSs in ${\cal D}_k$, with $\Lambda_{k,n}=\emptyset$ if $n > J_k$. Let $\Omega^{(l)}_{k,n}=\{a_1,a_2,\cdots,a_n\}$ represent the $l$-th reflection path in $\Lambda_{k,n}$, where $a_i$ denotes the index of the $i$-th intermediate IRS and it must hold that $d_{0,a_1} < d_{0,a_2} < \cdots <d_{0,a_n}$ for the signal to be reflected outwards from the BS to user $k$ in the downlink (or inwards in the uplink). Then, the effective MISO channel between the BS and user $k, k \in {\cal K}$, under the reflection path $\Omega^{(l)}_{k,n}$ is expressed as 
\begin{equation}\label{recvsig}
{\mv h}^{(l)}_{k,n}\!\!=\!\!
\begin{cases}
{\mv Q}_{0,a_1}\Big(\prod\limits_{1 \le i < n}{\mv \Phi}_{a_i}{\mv S}_{a_i,a_{i+1}}\Big){\mv \Phi}_{a_n}{\mv g}_{a_n,k} \!\!&\!\! {\text{if}\;}2 \!\le\! n \!\le\! J_k\\ 
{\mv Q}_{0,a_1}{\mv \Phi}_{a_1}{\mv g}_{a_1,k} \!\!&\!\! {\text{if}\;}n=1
\end{cases}.
\end{equation}
It is observed from (\ref{recvsig}) that as compared to the single-reflection link with $n=1$, the multi-reflection link with $n \ge 2$ suffers more severe multiplicative path loss over all its constituent $n+1$ link channels, i.e., ${\mv Q}_{0,a_1}$, ${\mv S}_{a_i,a_{i+1}}, 1 \le i < n$, and ${\mv g}_{a_n,k}$. For example, as shown in Fig.\,\ref{RefRegion}, the reflection path from the BS to user $k$ via IRSs 1, 3 and $J$ suffers a multiplicative path loss over its four constituent links. On the other hand, however, the multi-reflection link can reap a CPB gain by jointly designing the reflection of all $n$ IRSs in $\Omega^{(l)}_{k,n}$, which aims to compensate for the end-to-end path loss and thus yield a much stronger channel gain as compared to the direct BS-user channel ${\mv f}_k$. As will be shown later in Section \ref{MultiIRSOpt}, if the constituent links in $\Omega^{(l)}_{k,n}$ are all dominated by an LoS path, the CPB gain increases with the number of IRS reflections, $n$, and can be superior to the direct channel.

Based on the above, the effective MISO channel between the BS and user $k$ is given by the superposition of their direct channel and IRS-reflected channels with different number of IRS reflections, i.e.,
\begin{equation}\label{effeCh}
{\mv h}_k={\mv f}_k+\sum\limits_{n=1}^{J_k}\sum\limits_{l=1}^{\lvert \Lambda_{k,n} \rvert}{\mv h}^{(l)}_{k,n}, k \in {\cal K}.
\end{equation}
It is observed from (\ref{effeCh}) that the effective MISO channel ${\mv h}_k$ depends on all BS-IRS, inter-IRS and IRS-user channels in ${\cal D}_k$, as well as the passive beamforming of all IRSs in ${\cal D}_k$. 

\section{Reflection Design and Optimization for Double-IRS System}\label{db}
\begin{figure}[!t]
\centering
\includegraphics[width=3.2in]{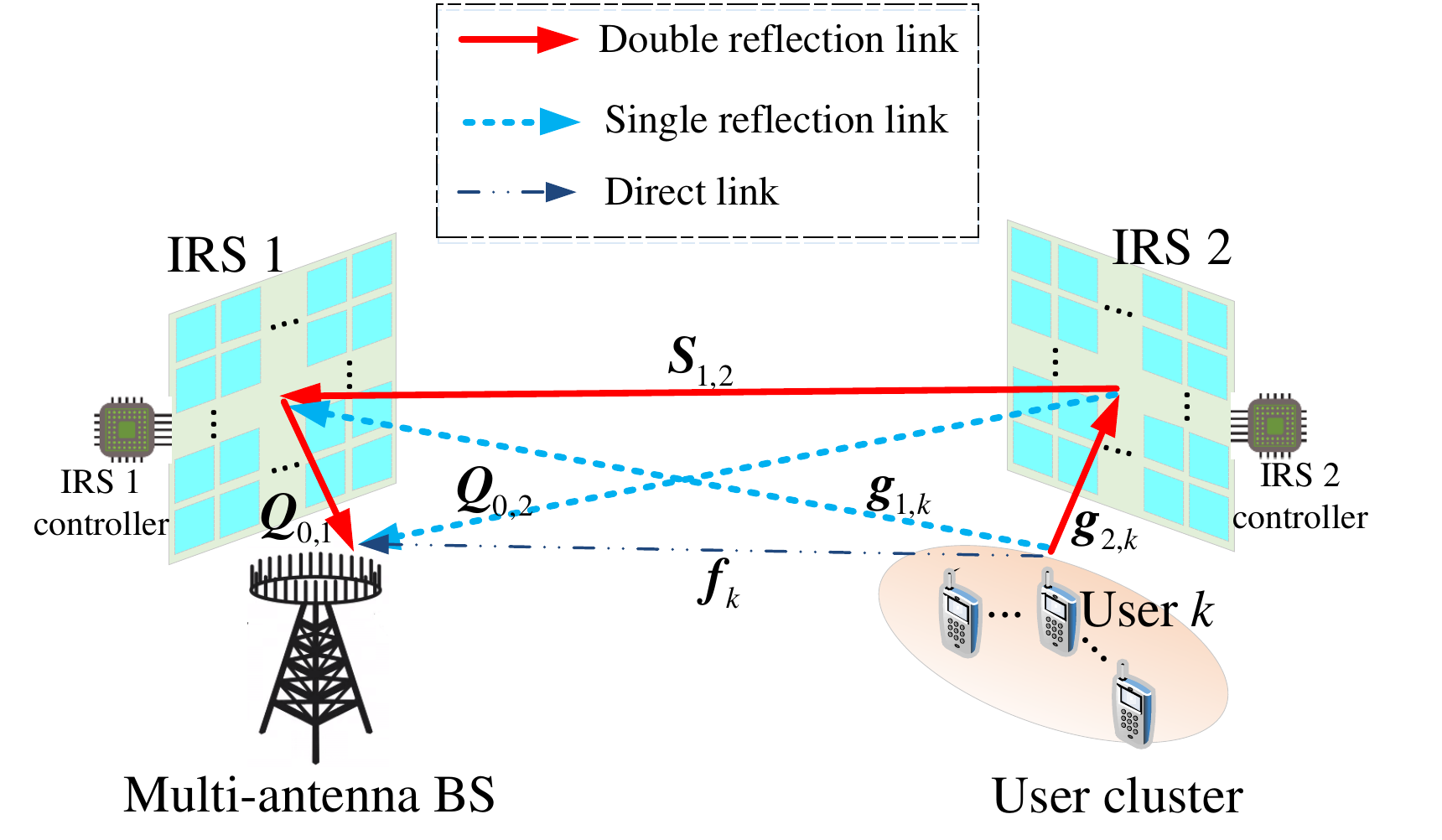}
\DeclareGraphicsExtensions.
\caption{A double-IRS cooperatively assisted multi-user communication system.}\label{system_2IRS}
\vspace{-12pt}
\end{figure}
In this section, we consider a double-IRS aided multi-user system, in which the communications between the multi-antenna BS and each cluster of users are assisted by two fixed IRSs that are separately deployed near the BS and the user cluster (for minimizing the path loss of their individual single-reflection links), thus referred to as the BS-side IRS and user-side IRS, respectively. For convenience, we consider one cluster of users, and refer to the BS-side IRS as IRS 1 and the user-side IRS as IRS 2, as shown in Fig.~\ref{system_2IRS}.
Note that this setup can be considered as a special case of the general multi-IRS system in Fig. 2 with $J=2$ and thus ${\cal D}_k={\cal J}=\{1,2\}, \forall k \in \cal K$.\footnote{More generally, this setup also incorporates the case of multiple BS-side IRSs and
	multiple user clusters each with a different user-side IRS, as we can treat these BS-side IRSs and user-side IRSs as two equivalent (aggregated) IRSs (i.e., IRS 1 and IRS 2, respectively) of a larger size accordingly.} Thus, in addition to the conventional BS-user direct links, the $K$ users can be effectively served by the BS through the two single- and one double-reflection links established by the two IRSs (i.e., IRS 1 and IRS 2). Under the above setup, the effective MISO channel between the BS and user $k, k \in {\cal K}$ in \eqref{effeCh} can be simplified as 
\begin{equation}\label{ch_model}
{\bm h}_k\!\!=\!\!\underbrace{{\bm f}_k}_{\text {Direct~link}}\!\!+ 
\underbrace{{\bm Q}_{0,1} {\bm \Phi}_1 {\bm g}_{1,k} \!+\! {\bm Q}_{0,2}{\bm \Phi}_2{\bm g}_{2,k}}_{\text {Single-reflection~links}}\!+\!
\underbrace{{\bm Q}_{0,1}{\bm \Phi}_1{\bm S}_{1,2}{\bm \Phi}_2 {\bm g}_{2,k}}_{\text {Double-reflection~link}}.
\end{equation}

\subsection{Double-IRS Reflection Optimization}\label{DB_Opt} 
\subsubsection{Double-IRS-Aided SISO System}\label{SU_SISO} 
First, we consider the single-user single-input-single-output
(SISO) setup, i.e., $K = 1$ and $N_B=1$. Moreover, we consider the worst-case propagation scenario where the direct BS-user, IRS 1-user, and BS-IRS 2 links are severely blocked and thus we only focus on the double-reflection (i.e., BS-IRS 1-IRS 2-user) link to draw essential and useful insights into the CPB design and the resultant received signal power scaling order in this case.
Accordingly, by denoting the baseband equivalent channel from the BS to IRS 1 as ${\bm q}_{0,1}\in {\mathbb C}^{M \times 1}$, the effective double-reflection SISO channel is modeled as (with the user index $k$ omitted for brevity)
\begin{equation}\label{DIRS_ch_model}
{h}={\bm q}_{0,1}^T{\bm \Phi}_1{\bm S}_{1,2}{\bm \Phi}_2 {\bm g}_{2}.
\end{equation}
Let us further consider the practical scenario where the two IRSs are properly deployed to achieve an LoS inter-IRS channel, which can be well approximated as ${\bm S}_{1,2}=\rho_{1,2} {\bm s}_1 {\bm s}_2^T $ with $\rho_{1,2}\in {\mathbb C}$ accounting for the complex-valued path gain, and ${\bm s}_1\in {\mathbb C}^{M \times 1}$ and ${\bm s}_2\in {\mathbb C}^{M \times 1}$ denoting the (transmit/receive) array response vectors at IRS 1 and IRS 2, respectively. In this case, the equivalent SISO channel in \eqref{DIRS_ch_model} can be rewritten as
\begin{align}
{h}&=\rho_{1,2}{\bm q}_{0,1}^T{\bm \Phi}_1{\bm s}_1 {\bm s}_2^T{\bm \Phi}_2 {\bm g}_{2}\nonumber\\
&=\rho_{1,2} \underbrace{{\bm q}_{0,1}^T {\rm diag}({\bm s}_1)}_{{\bm v}_{1}^T} {\bm \phi}_1 \cdot \underbrace{ {\bm s}_2^T {\rm diag}({\bm g}_{2}) }_{{\bm v}_{2}^T} {\bm \phi}_2.\label{DIRS_ch_model2}
\end{align}
Accordingly, the optimal passive beamforming vectors at IRS 1 and IRS 2 (denoted by ${\bm \phi}_1^{\star}$ and ${\bm \phi}_2^{\star}$, respectively) are designed as ${\bm \phi}_1^{\star}=e^{-j \angle \left({\bm v}_{1}\right)}$ and ${\bm \phi}_2^{\star}=e^{-j \angle \left({\bm v}_{2}\right)}$, so as to achieve the maximum channel power gain as \cite{han2020cooperative}
\begin{align}\label{DIRS_ch_gain}
|{h}^{\star}|^2=\max_{{\bm \phi}_1,{\bm \phi}_2}~|{h}|^2=|\rho_{1,2}|^2 \|{\bm v}_{1}\|_1^2  \cdot  \|{\bm v}_{2}  \|_1^2 \sim \frac{\beta^3 }{d_{\bm q} d_{\bm S}d_{\bm g}}M^4,
\end{align}
where $\beta$ denotes the path-loss at the reference distance of $1$ meter (m), and $d_{\bm q}$, $d_{\bm S}$, and $d_{\bm g}$ stand for the propagation distance for the BS-IRS 1, IRS 1-IRS 2, and IRS 2-user links, respectively. 
According to \eqref{DIRS_ch_gain}, a passive beamforming gain of order ${\cal O} (M^4)$ is achieved by properly aligning the passive beamforming directions of the two cooperative IRSs, which significantly outperforms the conventional single-IRS system with a passive beamforming gain of order ${\cal O} (M^2)$. This is because under the LoS inter-IRS channel condition, each of the two IRSs can achieve a passive beamforming gain of order ${\cal O} (M^2)$, together resulting in a much larger multiplicative CPB gain of order ${\cal O} (M^4)$.
However, one should keep in mind that as compared to the single-reflection link, the double-reflection link established by deploying two cooperative IRSs incurs higher product-distance path loss as $\beta^3$ shown in \eqref{DIRS_ch_gain}, and thus a sufficiently large value of $M$ is needed to compensate for such a high path loss to achieve an overall channel gain.
As such, depending on the number of IRS elements $M$ and inter-IRS channel condition (whether it is LoS dominant or not), there is a trade-off between the CPB gain and path loss in the selection of signal routing over the single- versus double-reflection links in the two IRSs' reflection design.

\begin{figure}[!t]
	\centering
	\includegraphics[width=2.2in]{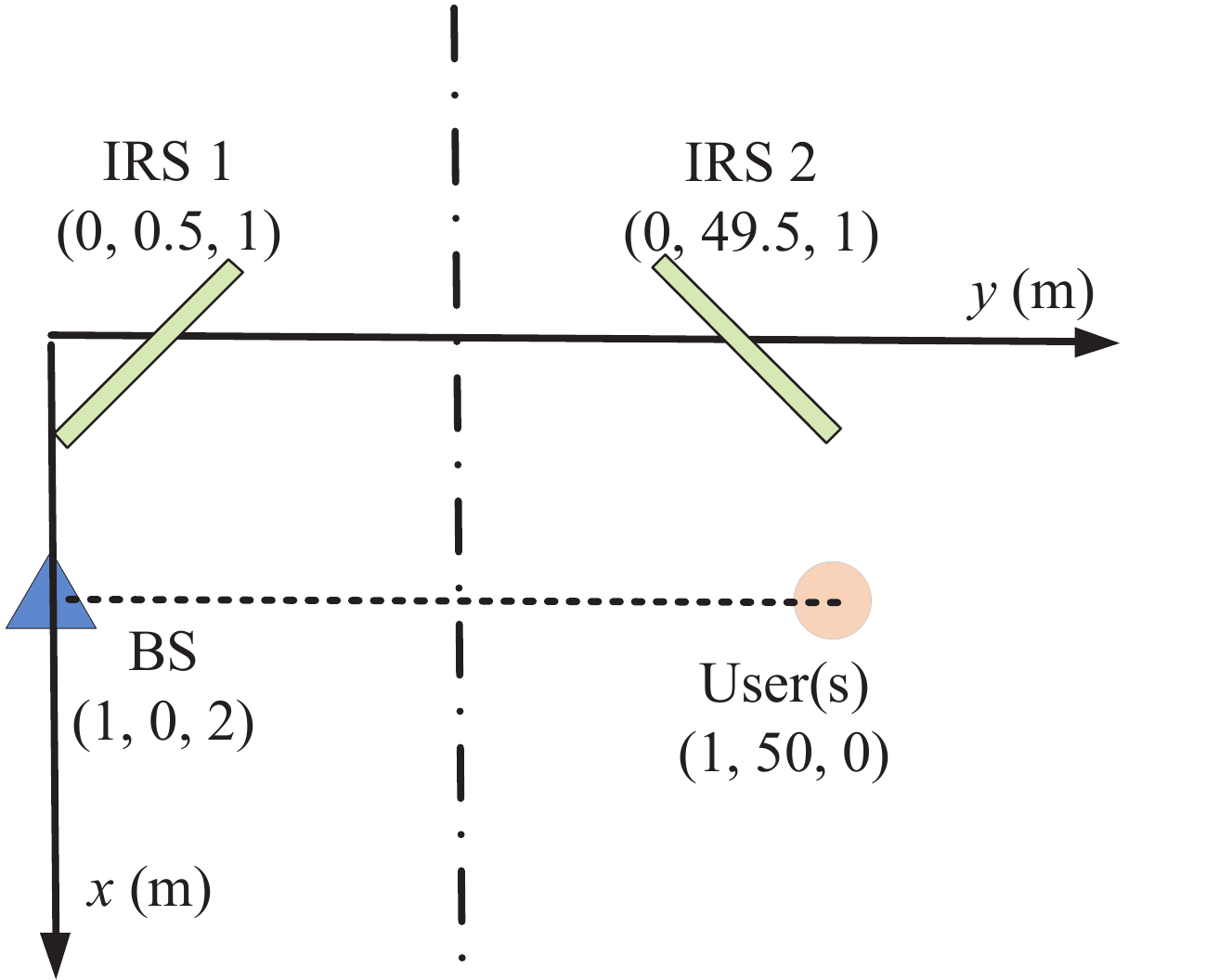}
	\DeclareGraphicsExtensions.
	\caption{Simulation setup of the double-IRS aided single-user system (Top view).}\label{system_2IRS_sim}
	\vspace{-3pt}
\end{figure}
\begin{figure}[!t]
	\centering
	\includegraphics[width=3.2in]{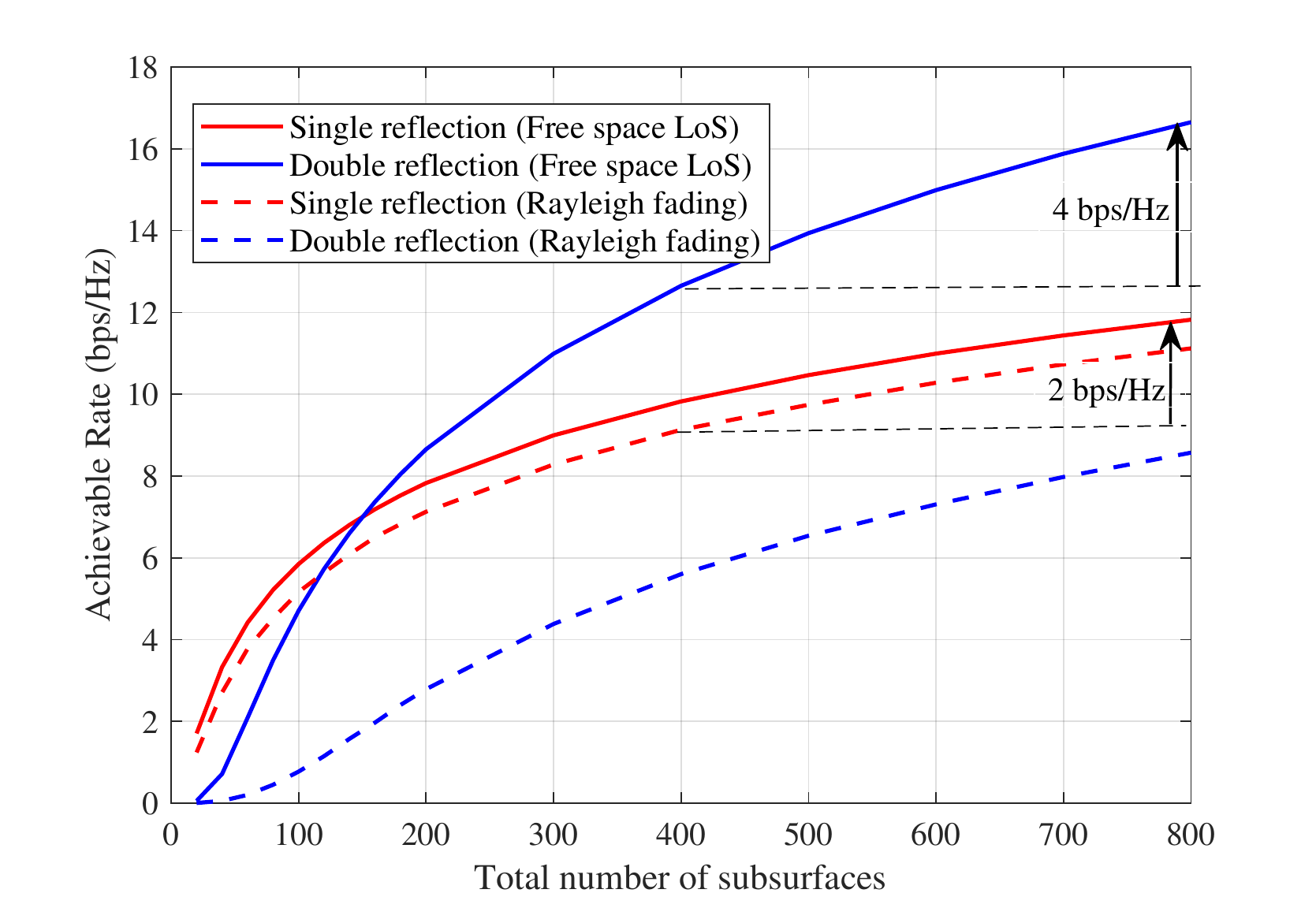}
	\DeclareGraphicsExtensions.
	\caption{Achievable rate versus the total number of IRS elements under different channel models.}\label{rate_element_numV2}
	\vspace{-3pt}
\end{figure}
To illustrate the received signal power scaling order and the performance trade-off between the passive beamforming gain and the path loss of single/double-reflection links, we consider a system setup shown in Fig.~\ref{system_2IRS_sim}, where the distance-dependent channel path loss is modeled as $\beta d_0^{-\alpha }$ with $d_0$ being the corresponding link distance and $\alpha$ accounting for the path loss exponent. Moreover, we set $\beta=-30$~dB, transmit power $P=0$~dBm, and noise power $\sigma^2=-90$~dBm. In Fig.~\ref{rate_element_numV2}, we plot the achievable rate against the total number of IRS reflecting elements by employing the two single-reflection (i.e., BS-IRS 1-user and BS-IRS 2-user) links versus the double-reflection (i.e., BS-IRS 1-IRS 2-user) link.  Moreover, we consider two different inter-IRS channel models, namely, free-space LoS channel and Rayleigh fading channel, with their path loss exponents set as $2$ and $2.5$, respectively.  In Fig.~\ref{rate_element_numV2}, one can observe that by doubling the total number of IRS elements from $400$ to $800$, the achievable rate of the double-reflection link under the free-space LoS channel increases about $\log_2(2^4) = 4$ bits-per-second-per-Hertz (bps/Hz); whereas that of the two single-reflection links only increases about $\log_2(2^2) = 2$ bps/Hz. This performance gap is expected since the two systems have different received signal power scaling orders, i.e., ${\cal O} (M^4)$ versus ${\cal O} (M^2)$ with increasing $M$ under the LoS channel condition. However, when the number of IRS elements is not large enough and/or the inter-IRS channel is Rayleigh fading or non-LoS (NLoS), the achievable rate of the double-reflection link is even lower than that of the two single-reflection links, due to its higher product-distance path loss and the fact that the CPB gain of ${\cal O}(M^4)$ over the inter-IRS LoS link is no more available.

\subsubsection{Double-IRS-Aided Single-User MISO System}\label{SU_MISO}
Next, we consider the single-user setup but with multiple antennas equipped at the BS, i.e., $K = 1$ and $N_B>1$. We also consider the worst-case scenario where the direct link between the BS and the user is severely blocked and thus can be ignored. In this case, the (active) transmit/receive beamforming at the BS needs to optimized jointly with the cooperative (passive) reflect beamforming at the two distributed IRSs. Let ${\bm w}^H\in {\mathbb C}^{1 \times N_B}$ denote the (active) transmit/receive beamforming at the BS and the resultant effective end-to-end channel between the BS and the user is given by
\begin{align}
{\bar h}&={\bm w}^H{\bm h}\nonumber\\
&= {\bm w}^H\!\!\left({\bm Q}_{0,1} {\bm \Phi}_1 {\bm g}_{1} \!+\! {\bm Q}_{0,2}{\bm \Phi}_2{\bm g}_{2}\!+\!
{\bm Q}_{0,1}{\bm \Phi}_1{\bm S}_{1,2}{\bm \Phi}_2 {\bm g}_{2}\right).\label{MISOch_model}
\end{align}
For any given ${\bm w}^H$, ${\bm \Phi}_1$, and ${\bm \Phi}_2$, we define the effective channel power gain for the single- and double-reflection links as $\gamma_s=|{\bm w}^H\left({\bm Q}_{0,1} {\bm \Phi}_1 {\bm g}_{1} + {\bm Q}_{0,2}{\bm \Phi}_2{\bm g}_{2}\right)|^2$ and $\gamma_d=|{\bm w}^H
{\bm Q}_{0,1}{\bm \Phi}_1{\bm S}_{1,2}{\bm \Phi}_2 {\bm g}_{2}|^2$, respectively.
It is expected that the channel gain maximization for the single- and double-reflection links, i.e., $\max\limits_{{\bm w}^H,{\bm \phi}_1,{\bm \phi}_2}~\gamma_s$ and $\max\limits_{{\bm w}^H,{\bm \phi}_1,{\bm \phi}_2}~\gamma_d$ will generally lead to different optimal joint active/passive beamforming designs of $\{{\bm w}^H,{\bm \phi}_1,{\bm \phi}_2\}$. In other words, we cannot find an optimal joint active/passive beamforming design that caters for the both single- and double-reflection links at the same time (unless for some special channel realizations).
As such, in order to maximize the effective end-to-end channel gain, i.e., $\max\limits_{{\bm w}^H,{\bm \phi}_1,{\bm \phi}_2}~|{\bar h}|^2$, we generally need to jointly design the active/passive beamforming {to strike an optimal balance between the channel gains} $\gamma_s$ and $\gamma_d$ over the single- and double-reflection links, respectively.\footnote{For simplicity, we can also design the joint active/passive beamforming design only catering for the dominant one of the single- and double-reflection links, which, however, is generally suboptimal.}
It is also worth pointing  out that the joint active/passive beamforming optimization is generally non-convex due to the unit-modulus
constraints of $\{{\bm \phi}_1,{\bm \phi}_2\}$ as well as the coupling between the active beamforming ${\bm w}^H$
and CPB $\{{\bm \phi}_1,{\bm \phi}_2\}$.
It was shown in \cite{zheng2021double} that the alternating optimization (AO) approach can be applied to solve this joint beamforming design problem efficiently by alternately optimizing the active beamforming at the BS and the CPB at the two IRSs in an iterative manner.
On the other hand, one can always design the two passive beamforming vectors as $e^{j\theta}{\bm \phi}_1$ and $e^{j\theta}{\bm \phi}_2$ for IRSs 1 and 2, respectively, to achieve coherent channel combining for the single- and double-reflection links, where  $\theta$ denotes a common phase shift applied to the two IRSs \cite{zheng2021double}. Specifically, by substituting $e^{j\theta}{\bm \phi}_1$ and $e^{j\theta}{\bm \phi}_2$ into \eqref{MISOch_model}, we obtain the corresponding channel gain as
\begin{align}
|{\bar h}|^2&=
\lvert e^{j\theta} \underbrace{{\bm w}^H\left({\bm Q}_{0,1} {\bm \Phi}_1 {\bm g}_{1} + {\bm Q}_{0,2}{\bm \Phi}_2{\bm g}_{2}\right)}_{a_s}\nonumber\\
&\qquad\quad+e^{j2\theta}  \underbrace{{\bm w}^H{\bm Q}_{0,1}{\bm \Phi}_1{\bm S}_{1,2}{\bm \Phi}_2 {\bm g}_{2}}_{a_d}\rvert^2\stackrel{(a)}{\le}\left(|a_s|+|a_d|\right)^2,\nonumber
\end{align}
where $(a)$ is due to the triangle inequality and the equality holds if and only if $\angle (e^{j\theta}a_s )=\angle (e^{j2\theta}a_d )$, which can be easily achieved with $\theta=\angle (a_s/a_d)$ for the coherent channel combination of the double- and single-reflection links.

\subsubsection{Double-IRS-aided Multi-user MISO System}\label{MU_MISO}
 Last, we consider the multi-user setup with multi-antenna BS, i.e., $K > 1$ and $N_B>1$, as shown in Fig.~\ref{system_2IRS}.
In this case, the joint active/passive beamforming optimization problem needs to cater for the multi-user channels with diverse QoS requirements/constraints, which thus are more involved and challenging to solve.
Moreover, the multi-antenna BS serves multiple users simultaneously over the same frequency band for the
downlink/uplink transmissions to achieve high spectral efficiency, which, however, generally suffers the multi-user co-channel interference. 

Let ${\bm H}=\left[{\bm h}_1,\ldots, {\bm h}_K\right]\in {\mathbb C}^{N_B \times K}$ denote the effective BS-user
channel matrix. Different from the single-user setup for reaping the passive beamforming gain as much as possible to maximize the effective channel gain (as shown in Sections \ref{SU_SISO} and \ref{SU_MISO}), the spatial multiplexing gain is more practically relevant under the multi-user setup so as to provide sufficient DoFs to mitigate the multi-user interference.
Note that the spatial multiplexing gain of multi-user systems critically depends on the rank of the multi-user effective channel ${\bm H}$.
Specifically, if ${\rm rank}({\bm H})\ge K$, we can fully mitigate the multi-user interference {by applying e.g., the zero-forcing (ZF)} transmit/receive beamforming at the BS. On the other hand, if ${\rm rank}({\bm H})<K$, we may not have enough DoFs to fully mitigate the multi-user interference and thus the system becomes {interference-limited}. 

\begin{figure}[!t]
	\centering
	\includegraphics[width=3.5in]{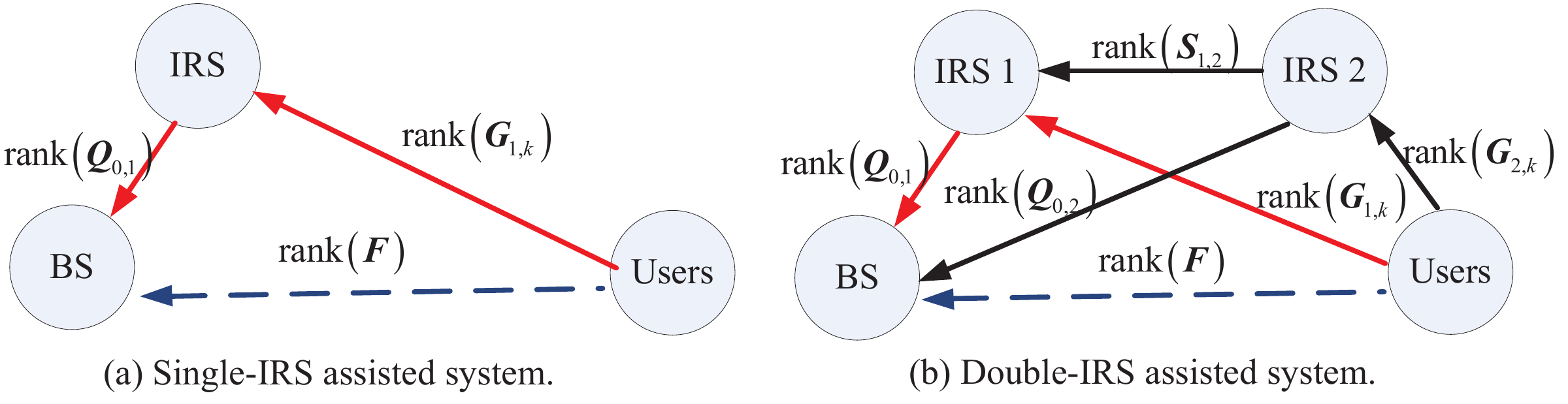}
	\DeclareGraphicsExtensions.
	\caption{Channel ranks for double- versus single-IRS assisted systems \cite{zheng2021double}.}\label{Maximum_flow}
	\vspace{-12pt}
\end{figure}
In contrast to traditional wireless communication systems where the MIMO channel rank is determined by the propagation environment, IRSs are able to add extra signal paths that are controllable to improve the MIMO channel rank condition for achieving higher spatial multiplexing gains. Moreover, as compared to the single-IRS aided system with one single-reflection link only (see Fig.~\ref{Maximum_flow}(a)), the double-IRS aided system significantly improves the path diversity by possessing not only two single-reflection links but also one additional double-reflection link (see Fig.~\ref{Maximum_flow}(b)), thus having the greater potential to bypass obstacles more effectively.
Moreover, it was shown in \cite{zheng2021double} that the double-IRS system has a higher channel rank than the single-IRS baseline in general, with a gain no less than $\min \left({\rm rank}({\bm G}_{2,k}),{\rm rank}({\bm Q}_{0,2}) \right)$, where ${\bm G}_{1,k}=\left[{\bm g}_{1,1},\ldots, {\bm g}_{1,K}\right]\in {\mathbb C}^{M \times K}$, ${\bm G}_{2,k}=\left[{\bm g}_{2,1},\ldots, {\bm g}_{2,K}\right]\in {\mathbb C}^{M \times K}$, and ${\bm F}=\left[{\bm f}_1,\ldots, {\bm f}_K\right]\in {\mathbb C}^{N_B \times K}$.
With a larger channel rank, the double-IRS system achieves higher spatial multiplexing gains for supporting more users, which leads to significantly improved multi-user signal-to-interference-plus-noise ratio (SINR)/rate performance over the single-IRS counterpart \cite{zheng2021double}. Under the multi-user setup, the authors in \cite{zheng2021double} extended the AO-based joint active/passive beamforming design by leveraging the semidefinite relaxation (SDR) and bisection methods to
solve the max-min SINR/rate problem efficiently for the double-IRS assisted system.

\begin{figure}[!t]
	\centering
	\includegraphics[width=3.2in]{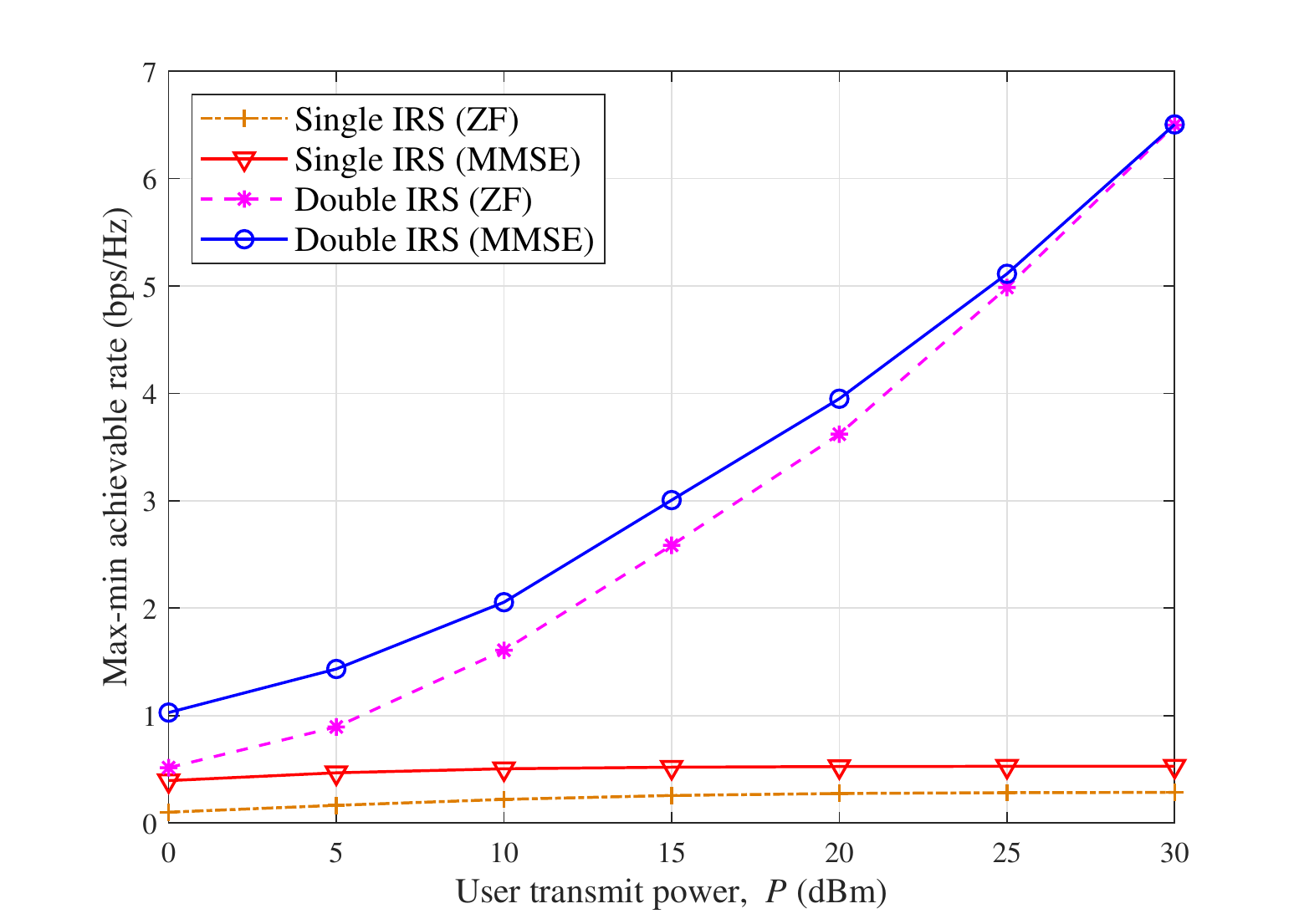}
	\DeclareGraphicsExtensions.
	\caption{Max-min achievable rate versus user transmit power $P$, with $K=5$.}\label{rate_SNR_MU}
	\vspace{-3pt}
\end{figure}

In Fig.~\ref{rate_SNR_MU}, we show the maximized minimum (max-min) achievable rate among all users of the double-IRS system against the conventional single-IRS baseline versus the user transmit power in the uplink communication, under the same setup as in Fig.~\ref{system_2IRS_sim} with $M=400$, $N_B=40$, and $K=5$. Under our simulation setup with ZF beamforming applied at the BS, the double-IRS system can achieve  ${\rm rank}({\bm H})=K$; while the multi-user effective channel of the single-IRS baseline is rank-deficient, which results in max-min rate saturation as the user transmit power increases. This is due to the lack of spatial DoFs to fully mitigate the multi-user interference and thus the single-IRS baseline becomes interference-limited.
In contrast, the double-IRS deployment achieves a much better channel rank condition and thus its max-min achievable rate keeps increasing with the user transmit power. Finally, for the system with a deficient channel rank, the MMSE receive beamforming generally achieves better performance than the ZF counterpart.

\subsection{Double-IRS Channel Estimation}\label{DIRS_CE}
For the case of fully-passive IRS 1 and IRS 2 (without any sensing ability) in the double-IRS aided system, it is practically infeasible to acquire the CSI between the two IRSs as well as that for them with the BS/users separately. Instead, only the cascaded CSI over the single- or double-reflection links can be estimated at one end point of the communication system, e.g., the BS that has more signal processing capability than users. In particular, it was shown in \cite{zheng2021uplink,zheng2021efficient} that the cascaded CSI over the single- and double-reflection links is sufficient for joint active/passive beamforming design without loss of optimality.
Accordingly, the ``separate" channel model in \eqref{ch_model} can be equivalently expressed in the ``cascaded" channel form as
\begin{align}
{\bm h}_k= {\bm f}_k&+
\underbrace{  {\bm H}_{0,1}  {\rm diag} ({\bm g}_{1,k}) }_{{\bm R}_{1,k}}{\bm \phi}_1+\underbrace{{\bm H}_{0,2}  {\rm diag} ({\bm g}_{2,k})}_{{\bm R}_{2,k}} {\bm \phi}_2 \nonumber\\
 &+\sum_{m=1}^{M} \underbrace{{\bm H}_{0,1}  {\rm diag} ({\bar{\bm s}}_{k,m})  }_{{\tilde{\bm S}}_{k,m}}{\bm \phi}_1{\phi}_{2,m} ,  \quad k \in {\cal K},\label{ch_model2}
\end{align}
where ${\bm R}_{1,k}\in {\mathbb C}^{M \times M}$ (${\bm R}_{2,k}\in {\mathbb C}^{M \times M}$) denotes the cascaded BS-IRS 1(IRS 2)-user $k$ channel for the two single-reflection links,
${\bar{\bm S}}_{k}\triangleq \left[{\bar{\bm s}}_{k,1},\ldots,{\bar{\bm s}}_{k,M} \right]={\bm S}_{1,2}  {\rm diag} ({\bm g}_{2,k})\in {\mathbb C}^{M \times M}$ denotes the cascaded IRS 1-IRS 2-user $k$ channel, and ${\tilde{\bm S}}_{k,m}$ stands for the cascaded BS-IRS 1-IRS 2-user $k$ channel associated with element $m$ of IRS 2, $m=1,\ldots,M$ for the double-reflection link.

\subsubsection{Double-IRS-Aided SISO System}\label{SU_SISO_CE} 
First, we consider the same single-user SISO setup as in Section~\ref{SU_SISO} with $K = 1$ and $N_B=1$. In this case, the effective double-reflection SISO channel in \eqref{DIRS_ch_model} can be rewritten as
\begin{equation}\label{DIRS_cascadedch_model}
{h}={\bm \phi}_1^T \underbrace{{\rm diag} ({\bm q}_{0,1}){\bm S}_{1,2}{\rm diag} ({\bm g}_{2})}_{\vec{\bm S}_{1,2}}  {\bm \phi}_2,
\end{equation}
where $\vec{\bm S}_{1,2}$ denotes the cascaded BS-IRS 1-IRS 2-user channel without applying any IRS phase shifts. Under the SISO setup with the general inter-IRS channel, it was shown in \cite{you2021wireless} that at least $M^2$ pilots symbols are required to estimate a total number of $M^2$ coefficients in the cascaded channel $\vec{\bm S}_{1,2}$.
On the other hand, if the inter-IRS channel is LoS, the equivalent SISO channel in \eqref{DIRS_ch_model} can be decomposed into two decoupled parts, ${\bm v}_{1}^T  {\bm \phi}_1$ and ${\bm v}_{2}^T  {\bm \phi}_2$, each corresponding to one of the two IRSs. In this case, for the passive beamforming design of $\{{\bm \phi}_1,{\bm \phi}_2\}$, we only need to acquire the two signature channel vectors ${\bm v}_{1}\in {\mathbb C}^{M \times 1}$ and ${\bm v}_{2}\in {\mathbb C}^{M \times 1}$, which can be estimated separately {by fixing one IRS's reflection while tuning the training reflection of the other IRS over time (similar to the single-IRS channel estimation proposed in \cite{OFDM_BX})}, thus leading to a reduced minimum training overhead of $2M$ pilot symbols.

\subsubsection{Double-IRS-aided Single-user MISO System}\label{SU_MISO_CE}
Next, we consider the single-user MISO system as in Section~\ref{SU_MISO}, i.e., $K = 1$ and $N_B>1$. In this case, even by ignoring the direct link\footnote{
	For the direct BS-user link, the BS can estimate it by using the conventional channel estimation method with the two IRSs both turned OFF.} in \eqref{ch_model2}, the total number of channel coefficients is prohibitively large for estimation, which consists of two parts:
\begin{itemize}
	\item The two single-reflection links with $2MN_B$ channel coefficients in ${\bm R}_{1}$ and ${\bm R}_{2}$;
	\item The double-reflection links with $ M^2N_B$ channel coefficients in $\{{\tilde{\bm S}}_{m}\}_{m=1}^M$.
\end{itemize}
From the above, the number of channel coefficients for the double-reflection link is of higher-order than that for the two single-reflection links, which makes the channel estimation problem more challenging for the double-IRS system in the MISO case than the SISO case. 

To tackle the above challenge, the authors in \cite{zheng2021uplink} exploited an interesting relationship between the single- and double-reflection channels to achieve efficient channel estimation with minimum training overhead. Specifically, it can be observed from \eqref{ch_model2} that the double-reflection channels $\{{\tilde{\bm S}}_{m}\}_{m=1}^M$ and the single-reflection channel ${\bm R}_{1}$ share the same (common) BS-IRS 1 channel, i.e., ${\bm H}_{0,1}$. 
As a result, if given the single-reflection channel ${\bm R}_{1}={\bm H}_{0,1}  {\rm diag} ({\bm g}_{1})$ as the reference
CSI, we can re-express ${\tilde{\bm S}}_{m}, m=1,2,\ldots,M$ in \eqref{ch_model2} as
\begin{align}\label{Sm}
{\tilde{\bm S}}_{m}={\bm H}_{0,1}  {\rm diag} ({\bar{\bm s}}_{m})=\underbrace{{\bm H}_{0,1} {\rm diag} ({\bm g}_{1})}_{{\bm R}_{1}} \cdot
\underbrace{{\rm diag} ({\bm g}_{1})^{-1} {\rm diag} ({\bar{\bm s}}_{m})}_{{\rm diag}({\bm a}_{m})},
\end{align}
with ${\bm a}_{m}={\rm diag} ({\bm g}_{1})^{-1} {\bar{\bm s}}_{m}$ accounting for the scaling vector, which implies that
the higher-dimensional double-reflection channels $\{{\tilde{\bm S}}_{m}\}_{m=1}^M$ can be effectively expressed as the lower-dimensional scaled versions of the single-reflection channel ${\bm R}_{1}$.
By substituting ${\tilde{\bm S}}_{m}$ of \eqref{Sm} into \eqref{ch_model2}, the channel model can be rewritten as
\begin{align}\label{double_ch}
{\bm h}= 
{\bm R}_{1}{\bm \phi}_1+{\bm R}_{2} {\bm \phi}_2
+\sum_{m=1}^{M} {\bm R}_{1}{\rm diag}({\bm a}_{m}){\bm \phi}_1{\phi}_{2,m}.
\end{align}
According to \eqref{double_ch}, it is sufficient to acquire the direct CSI ${\bm f}$, the cascaded CSI $\{{\bm R}_{1},{\bm R}_{2}\}$, and the scaling vectors $\{{\bm a}_{m}\}_{m=1}^M$ for the joint active/passive beamforming in the double-IRS system.
Based on the channel relationship shown in \eqref{Sm} and \eqref{double_ch}, a decoupled channel estimation scheme was proposed in \cite{zheng2021uplink} to successively estimate the CSI of ${\bm f}$, $\{{\bm R}_{1},{\bm R}_{2}\}$, and $\{{\bm a}_{m}\}_{m=1}^M$ in a decoupled manner with the ON/OFF IRS reflection design. Although this scheme  substantially reduces the training overhead by {circumventing the direct estimation of the high-dimensional double-reflection channel}, it generally suffers from not only residual interference due to imperfect signal cancellation but also considerable reflection power loss due to the ON/OFF reflection control of the two IRSs during the channel training.
Attentive to these issues, the authors in \cite{zheng2021efficient} further proposed to jointly estimate the CSI of the direct and single/double-reflection links with the always-ON training reflection to achieve the full-reflection power for improving the channel estimation accuracy significantly.
Moreover, it was shown in \cite{zheng2021uplink,zheng2021efficient} that the minimum training overhead is $2M+ \max\left\{M, \left\lceil \frac{M^2}{N_B}\right\rceil\right\}$, which decreases as the number of BS antennas $N_B$ increases until reaching its lower bound of $3M$.
As such, with a sufficiently large $N_B$ (say, $N_B\ge M$), the minimum training overhead of the double-IRS single-user system is comparable to that of the single-IRS counterpart, both of which linearly scale with the total number of IRS elements $M$ only.

\subsubsection{Double-IRS-aided Multi-user MISO System}\label{MU_MISO_CE}
Last, we consider the multi-user MISO system as in Section~\ref{MU_MISO}, i.e., $K >1$ and $N_B>1$.
In this multi-user case, the total number of channel coefficients in \eqref{ch_model2} is {$K$ times} that of the single-user case shown in Section~\ref{SU_MISO_CE}, which makes the multi-user channel estimation problem even more challenging.
As such, if a straightforward user-by-user successive channel estimation approach is applied, the total training overhead will be increased by {$K$ times} as compared to the single-user case.
To achieve more efficient multi-user channel estimation with affordable training overhead, the authors in \cite{zheng2021uplink,zheng2021efficient} further explored the channel relationship among multiple users.
Specifically, from \eqref{ch_model} and \eqref{ch_model2}, we can observe that all the users {{share the same (common)}} BS-IRS 1 (i.e., ${\bm H}_{0,1}$), BS-IRS 2 (i.e., ${\bm H}_{0,2}$), and IRS 1-IRS 2 (i.e., ${\bm S}_{1,2}$) links, {but have different} IRS 1-user (i.e., $\{{\bm g}_{1,k}\}_{k=1}^{K}$) and IRS 2-user (i.e., $\{ {\bm g}_{2,k}\}_{k=1}^{K}$) links in their respective single- and double-reflection channels (see Fig. \ref{system_2IRS}). 
Due to the above relationship, it was shown in \cite{zheng2021uplink,zheng2021efficient} that if given the cascaded CSI of an arbitrary user (say, user 1, which can be estimated as in the single-user case of Section~\ref{SU_MISO_CE}) as the reference CSI, we only need to estimate the low-dimensional scaling vectors ${\bm b}_{1,k}={\rm diag} ({\bm g}_{1,1})^{-1}{\bm g}_{1,k}$ and ${\bm b}_{2,k}={\rm diag} ({\bm g}_{2,1})^{-1}{\bm g}_{2,k}$ to acquire/recover the high-dimensional cascaded CSI of the remaining $K-1$ users with $k=2,\ldots,K$, thus substantially reducing the training overhead.
In particular, it was shown in \cite{zheng2021uplink,zheng2021efficient} that the additional training overhead for the remaining $K-1$ users can be as low as $\max\left\{K-1,\left\lceil \frac{2(K-1)M}{N_B}\right\rceil \right\}$, which further decreases with an increasing $N_B$ until reaching its lower bound of $K-1$.

\begin{figure}[!t]
	\centering
	\includegraphics[width=3.2in]{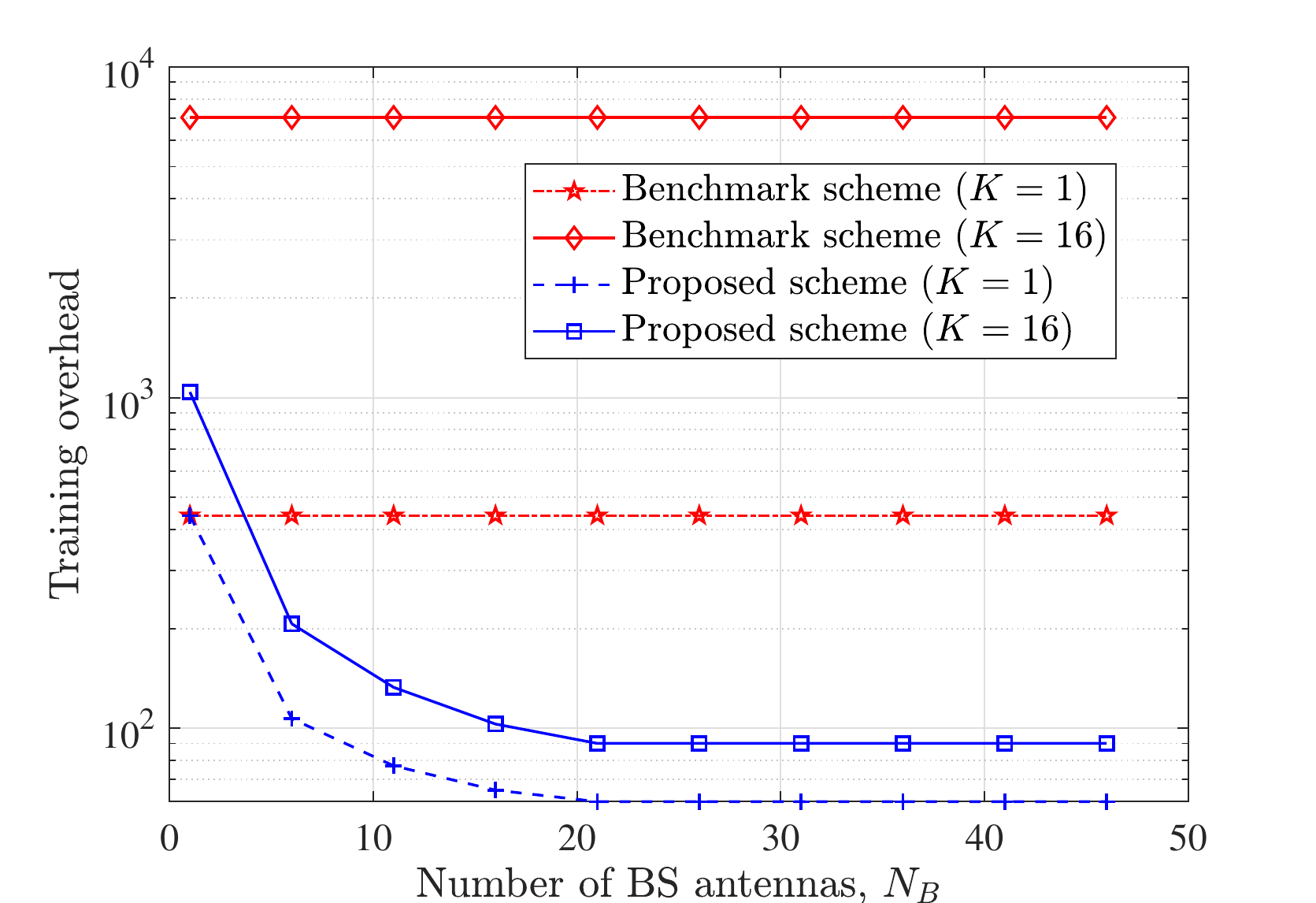}
	\DeclareGraphicsExtensions.
	\caption{Training overhead versus number of BS antennas $N_B$.}\label{overhead_vsN}
	\vspace{-12pt}
\end{figure}

In Fig.~\ref{overhead_vsN}, we show the required training overhead of the proposed scheme in \cite{zheng2021uplink,zheng2021efficient} against the benchmark scheme based on \cite{you2021wireless} versus the number of BS antennas, $N_B$. For the benchmark scheme based on \cite{you2021wireless}, the double-reflection channel is estimated at each BS antenna in parallel without exploiting the (common) channel relationship with the single-reflection channels, and the cascaded channels of $K$ users are separately estimated over consecutive time.
It is observed that for the proposed scheme in \cite{zheng2021uplink,zheng2021efficient}, the training overhead decreases dramatically with the increasing number of BS antennas $N_B$ for both the single- and multi-user cases, which is in sharp contrast to the benchmark scheme based on \cite{you2021wireless} where the training overhead is independent of $N_B$. As such, the training overhead can be substantially reduced by exploiting the (common) channel relationship between the single- and double-reflection channels as well as that among different users shown in Sections~\ref{SU_MISO_CE} and \ref{MU_MISO_CE}. Moreover, by increasing the number of antennas, the BS has more spatial observations for the uplink channel estimation given the same training time, which helps to reduce the training overhead in \cite{zheng2021uplink,zheng2021efficient}.

\subsection{Other Related Work and Future Direction}
In the last subsection, we discuss other related works/extensions on double-IRS aided communications and point out some promising directions to inspire future work.
\subsubsection{IRS Deployment/Placement}
Although this section mainly focuses on the double-IRS aided system by deploying one BS-side IRS and one user-side IRS to assist the communication for one user cluster only, it is worthy of extending the results to more general setups with multiple BS- and user-side IRSs to serve multiple distributed user clusters.
As compared to the conventional IRS deployment with IRS(s) placed either near the BS or user clusters only (for minimizing the product-distance path-loss), the new hybrid IRS deployment \cite{you2022deploy} with both BS- and user-side IRSs can not only attain  their complementary performance advantages, but also have more design flexibility in IRS reflection mode (single- or double reflection) selection for each user (cluster). In addition, deploying both BS- and user-side IRSs in the network can significantly increase the path diversity with both single- and double-reflection links, which makes the network {more {\it resilient} to the signal blockage}. Despite its great potential, this new IRS deployment also faces new and unique challenges that need to be addressed carefully.
For example, from the deployment perspective, how to optimally place and allocate the BS- and user-side IRSs in a given  wireless network for {balancing different reflection channel gains} as well as achieving more effective control of the radio environment is an important but also challenging problem to investigate.
Moreover, it is expected that the LoS availability of inter-IRS channels also highly depends on the deployment of BS- and user-side IRSs (e.g., {the direction facing} between the BS- and user-side IRSs), which is worthy of further investigation in the future.

\subsubsection{IRS-User Association}
Under the deployment of multiple BS- and user-side IRSs to serve multiple distributed user clusters, how to properly associate different IRSs with different user clusters is a critical problem for optimizing the network performance. In particular, {depending on the communication requirements and channel conditions (e.g., LoS link availability)}, each user or user cluster can be associated with no IRS, some BS-side IRSs only, some user-side IRSs only, or both the BS- and user-side IRSs.
Generally speaking, different IRS-user associations will lead to different transmission modes over various communication links, such as direct links, single-/double-reflection links or their various combinations with {different CSI requirements}.
Moreover, the IRS-user association problem is highly coupled with other communication designs/setups, such as the active/passive beamforming design at the BS/IRSs, the available CSI among different links, the users' mobility, etc., which deserves an in-depth study in the future.
\subsubsection{Resource Allocation and User Scheduling}
For the multi-IRS aided wireless network, resource allocation is an important topic that aims to achieve efficient utilization of various radio resources in the system, such as power, bandwidth, time, antennas, as well as IRS elements. 
However, due to the hardware limitation of IRS passive reflection that can only be made time-selective but not frequency-selective \cite{Zheng2020IRSNOMA}, dynamic resource allocations and user scheduling policies become more challenging to optimize as compared to traditional wireless systems.
Moreover, under the hybrid IRS deployment \cite{you2022deploy} with both BS- and user-side IRSs, one open challenge lies in how to jointly allocate a large number of reflecting elements over the BS- and user-side IRSs as well as time-frequency resources for different users to meet their diversified QoS requirements.
Moreover, if users require to associate with more than one BS- and user-side IRSs, the coordination between different IRSs is generally needed for efficient user scheduling over various resources.
As of now, the study on resource allocation and user scheduling under the hybrid IRS deployment is still at an initial stage and thus deserves more research efforts in the future.

\subsubsection{Other Extensions}
This section mainly focuses on the time-division duplexing (TDD) based double-IRS systems to facilitate the exposition of the uplink-downlink channel reciprocity, while the design in terms of reflection optimization and channel estimation in frequency-division duplexing (FDD) based systems is also practically important as well as challenging, which deserves further studies in the future.
Besides, the results presented in this section are mainly under the narrowband setup over flat-fading channels, while their extensions to the broadband systems with frequency-selective fading channels remain largely open and thus are worth further investigating. 
Note that in contrast to the narrowband system, the cascaded channels in the broadband system result in the {convolution of the IRS-aided channels}, which is even more sophisticated for the double-reflection link due to the higher-order channel convolution and thus makes both reflection optimization and channel estimation much more challenging than the single-IRS reflection link.
Furthermore, existing works on double-IRS systems mainly focus on the reflection design under perfect CSI, which are generally difficult to realize in practice. As such, more practical and robust designs based on imperfect CSI or partial/implicit CSI (e.g., codebook-based beam training without explicit channel estimation) also deserve further investigation in the future.

\section{Reflection Design and Optimization for Multi-IRS System}\label{MultiIRSOpt}
In this section, we consider the IRS reflection design and optimization problems arising in a general multi-IRS system with multiple signal reflections. For notational simplicity, we assume $\cup_{k \in \cal K}{\cal D}_k={\cal J}$, such that all IRSs are involved in the reflection design and channel estimation. 

\subsection{LoS-Dominant Channel Model}\label{SysMod_MI}
As shown in (\ref{effeCh}), for each user $k, k \in \cal K$, its effective MISO channel with the BS is determined by all BS-IRS, inter-IRS and IRS-user channels in ${\cal D}_k$. Thus, if all CSI in (\ref{effeCh}) is available, AO-based algorithm as in Section \ref{DB_Opt} can be applied to alternately optimize the active and passive beamforming at the BS and each individual IRS, respectively, until the convergence is reached. In the special case that there only exists a single and common reflection path from the BS to all users, this joint beamforming design problem can also be solved by exploiting the deep reinforcement learning based approach\cite{huang2021multi} or block minorization-maximization algorithm\cite{zhang2021weighted}. However, in the above works, the channel knowledge on all links in ${\cal D}_k, k \in \cal K$ needs to be perfectly known, thus incurring excessively high overhead for channel training and estimation. In particular, for a multi-reflection link involving more than two IRSs, it is difficult to extend the existing cascaded channel estimation methods proposed for single- and double-reflection links (see Section \ref{DIRS_CE}) to estimate the essential CSI required for IRS reflection optimization. Fortunately, by densely and properly deploying IRSs in the network, LoS propagation can be achieved for many BS-IRS, inter-IRS, and IRS-user links to enable efficient CPB over them, which also helps simplify the channel estimation problem. Hence, in this section, we focus on the LoS links in the system and solve their associated IRS reflection design and channel estimation problems. As a result, all NLoS links will be treated as part of environment scattering, which generally have a marginal effect on the system performance\cite{mei2021mbmh,mei2021distributed}. 

To describe the LoS availability between any two nodes $i$ (BS/IRS) and $j$ (IRS/user) in our considered system, we use a set of binary LoS condition indicators, denoted as $u_{i,j} \in \{0,1\}$. Based on Section \ref{SysModel}, we set $u_{i,j}=1$ if and only if the following three conditions are satisfied: 1) there is an LoS path between nodes $i$ and $j$; 2) $d_{0,j}>d_{0,i}$ (for reflecting signal outwards from the BS in the downlink case) except that node $j$ is the user; and 3) effective signal reflection can be achieved between nodes $i$ and $j$ subject to the 180$^\circ$ half-space reflection by each IRS. Otherwise, we set $u_{i,j}=0$. In addition, if an IRS $j, j \in \cal J$ is outside the effective IRS region ${\cal D}_k$ of user $k, k \in \cal K$, we also set $u_{j,J+k}=0$. Since the locations of the BS and all IRSs are fixed, the BS-IRS and inter-IRS (LoS) channels usually remain constant over a long period. As such, the LoS indicators between any two nodes (IRS/BS) can be obtained offline by leveraging the coordination among the BS and different IRS controllers based on existing LoS identification approaches (e.g., \cite{xiao2014non}). On the other hand, the LoS availability between each IRS in ${\cal D}_k$ and user $k, k \in \cal K$ should be identified via their real-time (online) coordination due to user mobility. To ensure the accuracy of such LoS identification, each IRS controller $j$ can be placed closely to the reference point of IRS $j$, such that its large-scale path loss with another node (BS/IRS/user) is approximately equal to that of the reflecting elements of IRS $j$. 

Based on the identified LoS conditions, we define $\tilde\Lambda_{k,n}$ as the set of all $n$-reflection ($n \ge 1$) LoS paths from the BS to user $k$ via the IRSs in ${\cal D}_k$, with $\tilde\Lambda_{k,n} \subseteq \Lambda_{k,n}$. Then, the effective MISO channels in (\ref{effeCh}) can be simplified as
\begin{equation}\label{LoSCh}
{\mv h}_k = {\mv f}_k+\sum\limits_{n=1}^{J_k}\sum\limits_{l=1}^{\lvert \tilde\Lambda_{k,n} \rvert}{\mv h}^{(l)}_{k,n}, k \in {\cal K},
\end{equation}
where the reflection paths comprising the NLoS propagation (i.e., the paths in the sets $\Lambda_{k,n} \backslash \tilde\Lambda_{k,n}$ for all $k$ and $n$) are omitted for convenience. In fact, it has been shown in \cite{mei2021massive} that the maximum strength of each of these paths can be much lower than that of an LoS end-to-end path even taking into account its more severe path loss, thanks to the more pronounced CPB gain over multiple LoS links.

Nonetheless, based on either (\ref{effeCh}) or (\ref{LoSCh}), it is generally difficult to obtain the optimal BS and IRS beamforming solution, as they are intricately coupled with each other in different reflection paths, although similar AO-based algorithm as in Section \ref{DB_Opt} can be applied to attain suboptimal solutions assuming perfect CSI. To further simplify the beamforming design (and ease the channel estimation to be shown in Section \ref{training}), we aim to select the best reflection path among all paths in $\tilde\Lambda_{k,n}, 1 \le n \le J_k$ for each user $k, k \in \cal K$, thanks to the abundant paths available with dense IRS deployment and then investigate the joint BS/IRS (active/passive) beam routing design with optimal IRS/path selection accordingly. This new beam routing problem is uniquely challenging for the multi-IRS system, although its preliminary form has been considered in Section \ref{db} for the double-IRS system in terms of single-/double-reflection link selection. 

\subsection{Multi-Reflection Beam Routing Optimization}\label{MUMR}
In this subsection, we present how to solve the joint active/passive beam routing optimization problem for the single-user and multi-user setups, respectively.  

{\it 1) Single User:} In the case of single user (e.g., user 1) in the snapshot of interest, we consider an arbitrary LoS reflection path $\Omega=\{a_1,a_2,\cdots,a_n\}$ from the BS to user 1 via $n$ IRSs in ${\cal J}$. Let $a_0=0$ and $a_{n+1}=J+1$, corresponding to the BS and user 1, respectively. Then, it must hold that $u_{a_i,a_{i+1}}=1, 0 \le i \le n$. To reveal valuable insights, we first consider the worst-case scenario with the BS-user channel severely blocked and thus assumed negligible, i.e., ${\mv f}_1 \approx {\bf 0}$. Thus, the effective MISO channel between the BS and the user under this LoS reflection path is given by
\begin{equation}\label{MRLink}
{\mv h}_{0,J+1}(\Omega)={\mv Q}_{0,a_1}\prod\limits_{1 \le i < n}\!\!\!\left({\mv \Phi}_{a_i}{\mv S}_{a_i,a_{i+1}}\right){\mv \Phi}_{a_n}{\mv g}_{a_n,J+1},
\end{equation}
and the corresponding end-to-end SISO BS-user channel is
\begin{equation}\label{LoSrecvsig}
h_{0,J+1}(\Omega)={\mv w}^H_B{\mv h}_{0,J+1}(\Omega), 
\end{equation}
where $\mv w_B^H \in {\mathbb C}^{1 \times N_B}$ denotes the (active) beamforming vector at the BS to serve user 1, with $\lvert \mv w_B \rvert^2=1$. 

Similarly as in Section \ref{MU_MISO}, if an LoS-dominant channel can be achieved between any two consecutive nodes in $\Omega$, we have ${\mv Q}_{0,a_1} \!\approx\! {\sqrt \beta}d^{-1}_{0,a_1}{\tilde{\mv q}}_{a_1,1}{\tilde{\mv q}}^H_{a_1,2}$, ${\mv S}_{a_i,a_{i+1}} \approx {\sqrt \beta}{d^{-1}_{a_i,a_{i+1}}}{\tilde{\mv s}}_{a_i,a_{i+1},1}{\tilde{\mv s}}^H_{a_i,a_{i+1},2}, 1 \le i <n$, and ${\mv g}_{a_n,J+1} = {\sqrt \beta}{d^{-1}_{a_n,J+1}}{\tilde{\mv g}}_{a_n,J+1}$, where ${\tilde{\mv q}}_{a_1,1}$, ${\tilde{\mv q}}_{a_1,2}$, ${\tilde{\mv s}}_{a_i,a_{i+1},1}$, ${\tilde{\mv s}}_{a_i,a_{i+1},2}$, and ${\tilde{\mv g}}_{a_n,J+1}$ are array responses and have unit-modulus entries. Then, by substituting them into (\ref{LoSrecvsig}) and rearranging the terms, it can be shown that the optimal IRS passive beamforming vectors that maximize the BS-user channel power gain, $\lvert h_{0,J+1}(\Omega) \rvert^2$, are given by\cite{mei2021cooperative} 
\begin{equation}\label{pbv}
{\mv \theta}_{a_i}=
	\begin{cases}
	{\rm{diag}}({\tilde{\mv q}}_{a_1,2})\cdot \tilde{\mv s}^*_{a_1,a_2,1}  &{\text{if}}\;i=1\\
	{\rm{diag}}({\tilde{\mv s}}_{a_{i-1},a_i,2}) \cdot \tilde{\mv s}^*_{a_i,a_{i+1},1} &{\text{if}}\;2 \le i \le n-1\\
	{\rm{diag}}({\tilde{\mv s}}_{a_{n-1},a_n,2}) \cdot \tilde{\mv g}^*_{a_n,J+1},&{\text{otherwise}}
	\end{cases}
\end{equation}
while the optimal BS active beamforming is given by the maximum ratio transmission (MRT) based on its array response with the first reflecting IRS, IRS $a_1$, i.e.,
\begin{equation}\label{abv}
	{\mv w}_B={\tilde{\mv q}}_{a_1,1}/\lVert {\tilde{\mv q}}_{a_1,1} \rVert={\tilde{\mv q}}_{a_1,1}/\sqrt{N}.
\end{equation}

As such, the maximum BS-user channel power gain under the path $\Omega$ is expressed as
\begin{equation}\label{eq1}
\lvert h_{0,J+1}(\Omega) \rvert^2=M^{2n}N_B\beta^{n+1}\prod\limits_{i=0}^{n}d^{-2}_{a_i,a_{i+1}}.
\end{equation}

It is observed from (\ref{eq1}) that in addition to the conventional active beamforming gain of $N_B$ by the BS, the $n$ IRSs in $\Omega$ also provides a CPB gain of $M^{2n}$, which increases exponentially with the number of IRS reflections $n$ under LoS channels. Nonetheless, the multi-reflection path $\Omega$ also suffers a multiplicative path loss, which also increases with $n$ in (\ref{eq1}). As such, there exists a fundamental trade-off between maximizing the CPB gain and minimizing the end-to-end path-loss in the optimal beam routing design of $\Omega$, so as to maximize $\lvert h_{0,J+1}(\Omega) \rvert^2$. Note that in addition to $\Omega$, there in general exist other signal paths (i.e., the paths in the sets $\Lambda_{1,n} \backslash \{\Omega\}$ for all $n$) from the BS to user 1. However, the strength of these (randomly) scattered paths is practically much lower than that of $\Omega$ under the optimized beamforming designs in (\ref{pbv}) and (\ref{abv}) based on the path $\Omega$ and thus can be ignored in the beam routing design\cite{mei2021mbmh,mei2021distributed}.

Given the maximum effective channel gain for each reflection path $\Omega$ in (\ref{eq1}), the optimal reflection path can be obtained by solving the following beam routing problem, i.e.,
\begin{equation}\label{op1}
{\text{(P1)}} \mathop {\max}\limits_{\Omega}\; \lvert h_{0,J+1}(\Omega) \rvert^2,\;\;\text{s.t.}\;\;u_{a_i,a_{i+1}}=1, 0 \le i \le n.
\end{equation}
Although (P1) is a combinatorial optimization problem, it can be optimally solved by resorting to graph theory. Specifically, we can construct an {\it LoS graph} for all the nodes and their wireless links in the system, denoted as $G_L = (V_L,E_L)$. The vertex set $V_L$ is given by $V_L={\cal J} \cup \{0,J+1\}$, and the edge set $E_L$ is defined as $E_L=\{(i,j)| u_{i,j} = 1\}$, i.e., there exists an edge from vertex $i$ to vertex $j$ if and only if there is an effective LoS link between them (subject to the three conditions specified in Section \ref{SysMod_MI}). It follows that any cascaded LoS reflection path from the BS to user 1 corresponds to a path from node 0 to node $J+1$ in $G_L$. Then, by properly assigning a weight to each edge in $G_L$, (P1) can be equivalently recast as the shortest path problem that aims to find the shortest path from node 0 to node $J+1$ in $G_L$, to which the classical Bellman-Ford algorithm can be applied to obtain the optimal solution efficiently\cite{mei2021cooperative}. 

On the other hand, if the direct BS-user channel ${\mv f}_1$ cannot be ignored, it can be shown that the maximum end-to-end effective channel power in (\ref{eq1}) becomes
\begin{align}
\lvert h_{0,J+1}(\Omega) \rvert^2=\lVert {\mv f}_1 \rVert^2 &+ M^{2n}N_B\beta^{n+1}\prod\limits_{i=0}^{n}d^{-2}_{a_i,a_{i+1}}+\nonumber\\
&2M^n\beta^{\frac{n+1}{2}}{\prod\limits_{i=0}^{n}d^{-1}_{a_i,a_{i+1}}}\lvert\tilde{\mv q}_{a_1,1}^H{\mv f}_1 \rvert.\label{eq1.dr}
\end{align}
Note that to achieve (\ref{eq1.dr}), the optimal BS and IRS beamforming is given by the MRT based on the effective channel and (\ref{pbv}), respectively, and a common phase shift is further added to one of the reflecting IRSs to achieve the coherent combining of the signals over $\Omega$ and the direct link at the user receiver. Hence, as shown in (\ref{eq1.dr}), the effective channel gain consists of three terms, corresponding to the strength of the direct BS-user link, selected multi-reflection link, as well as their crosstalk after the coherent combining at the user receiver. However, due to the presence of the third term, the resulting beam routing problem cannot be solved by applying the same approach as in (P1) if $N_B > 1$. Instead, we can only traverse all feasible reflection paths (or the paths from vertex 0 to vertex $J+1$ in $G_L$) and select the one that achieves the largest value of (\ref{eq1.dr}). Some existing algorithms in graph theory, e.g., depth-first search, can be utilized to facilitate this path traversal.

{\it 2) Multiple Users:} The multi-beam routing design for multiple users is more challenging as compared to that for a single user due to the following reasons. First, the multi-beam routing design may be coupled among different users and an IRS may be involved in the reflection paths of multiple users. Second, how to mitigate the inter-user interference in beam routing design is also a new and important issue to be tackled. In general, the previously proposed optimal single-beam routing design for a single user is not applicable to the case with multiple beams/users. Next, we present how to tackle this challenging problem in the multi-user case.

First, if all direct BS-user links are severely blocked, i.e., ${\mv f}_k \approx {\bf 0}, k \in \cal K$, then the inter-user interference over them can be ignored. As such, it suffices to mitigate the inter-user interference over the selected multi-reflection links between the BS and users. To this end, we can introduce a new type of {\it path separation} constraints among different users, where the IRSs selected for different users or paths are sufficiently separated, such that they do not have direct LoS links with each other, so as to mitigate the inter-user/path interference due to their undesired scattering. Moreover, this also helps simplify the BS and IRS beamforming design as it can be decoupled among different users and thus optimized similarly as in the single-user setup. 

Specifically, let $\Omega^{(k)}=\{a^{(k)}_1,a^{(k)}_2,\cdots,a^{(k)}_{n_k}\}, k \in \cal K$ denote the reflection path from the BS to user $k$, where $n_k\, (\ge 1)$ and $a^{(k)}_i \in \cal J$ denote the number of IRSs in $\Omega^{(k)}$ and the index of the $i$-th IRS with $1 \le i \le n_k$, respectively. Denote by ${\mv w}^H_{B,k} \in {\mathbb C}^{1 \times N_B}$ the BS's active beamforming to serve user $k, k \in \cal K$. Similarly as in the case of a single user, the maximum end-to-end channel power gain from the BS to user $k$ under the reflection path $\Omega^{(k)}$ is obtained as
\begin{equation}\label{eq2}
\lvert h_{0,J+k}(\Omega^{(k)}) \rvert^2\!=\!\!{N_BM^{2n_k}\beta^{n_k+1}}{\prod\limits_{i=0}^{n_k}d^{-2}_{a^{(k)}_i,a^{(k)}_{i+1}}}, k \in {\cal K}.
\end{equation}
Note that to achieve (\ref{eq2}), the MRT precoding should be applied at the BS to serve the $K$ users, based on its array responses with the $K$ selected first reflecting IRSs, i.e., ${\mv w}_{B,k} = {{\tilde{\mv q}}_{a^{(k)}_1,1}}/{\lVert {\tilde{\mv q}}_{a^{(k)}_1,1} \rVert}$, which, however, may result in inter-user interference over the BS-IRS links in the $K$ reflection paths. Fortunately, as the BS-IRS channels remain approximately static, all first reflecting IRSs can be properly deployed to avoid their mutual interference from the BS, even with the suboptimal MRT-based precoding. For example, if they are sufficiently separated in the angular domain and $N_B$ is sufficiently large (e.g., in massive MIMO), the inter-user interference over the BS-IRS links can be approximately nulled with MRT precoding\cite{mei2021mbmh,mei2021massive}. Since the path separation constraints ensure that the scattered inter-user interference via the subsequent inter-IRS and IRS-user links is well mitigated, each user is approximately free of inter-user interference, while achieving the maximum cascaded LoS channel gain with the BS as shown in (\ref{eq2}).

Given the maximum end-to-end channel gains in (\ref{eq2}) for the $K$ users, we can jointly optimize the reflection paths $\Omega^{(k)}, k \in \cal K$ to achieve their optimal performance in a fair manner, e.g., by maximizing their minimum achievable SINRs. Due to the well mitigated inter-user interference at each user's receiver, this is equivalent to maximizing the minimum BS-user LoS channel power gain, i.e., $\mathop {\min}\nolimits_{k \in \cal K} \lvert h_{0,J+k}(\Omega^{(k)})\rvert^2$. Thus, the multi-beam routing problem is formulated as
\begin{align}
{\text{(P2)}} &\mathop {\max}\limits_{\{\Omega^{(k)}\}_{k \in \cal K}}\;\mathop {\min}\limits_{k \in \cal K}\;\; \lvert h_{0,J+k}(\Omega^{(k)}) \rvert^2,\nonumber\\
\text{s.t.}\;\;&u_{a^{(k)}_i,a^{(k)}_{i+1}}=1, 1 \le i \le n_k, k \in {\cal K},\label{op3a}\\
&u_{a^{(k)}_i,a^{(k')}_{i'}}=0, \;a^{(k)}_i \ne a^{(k')}_{i'}, \forall i,i' \ne 0, k,k' \in {\cal K}, k \ne k'.\label{op3b}
\end{align}
Note that in (P2), the constraints in (\ref{op3a}) ensure that each of the BS-IRS, inter-IRS and IRS-user links in $\Omega^{(k)}$ consists of an effective LoS link, while those in (\ref{op3b}) ensure that there is no direct LoS link between any two nodes belonging to different reflection paths (except the common node $0$ or the BS). It has been shown in \cite{mei2021mbmh} that these constraints suffice to mitigate the interference power at each user receiver below the typical noise power under practical wireless system setups.

As compared to (P1), (P2) turns out to be a more challenging problem to solve due to the new path separation constraint (\ref{op3b}). In addition to the trade-off between the CPB gain and the end-to-end path loss in the beam routing design for each of the $K$ users, there also exists another trade-off in balancing all $\lvert h_{0,J+k}(\Omega^{(k)}) \rvert$'s for different users in $\cal K$. Specifically, due to the finite number of IRSs and LoS paths in the system as well as the path separation constraints in (\ref{op3b}), maximizing the channel gain for one user generally reduces the number of feasible reflection paths for the other users. As such, the multi-beam routing should be properly designed to reconcile the above two trade-offs at the same time. In fact, it has been shown in \cite{mei2021mbmh,mei2021massive} that (P2) is equivalent to an NP-complete problem by recasting it as a graph-theoretic problem based on the LoS graph of the system. To tackle this challenging problem, an efficient recursive algorithm was proposed in \cite{mei2021mbmh,mei2021massive} to partially enumerate the feasible reflection paths, which flexibly balances the performance and complexity. 

However, if the direct BS-user channels cannot be ignored, the path separation constraints may not be sufficient to mitigate the inter-user interference. In this case, the active precoding at the BS needs to take into account the channels between the BS and all users (in addition to all first reflecting IRSs) to eliminate the inter-user interference over both the direct BS-user and BS-IRS links, jointly with the path separation to mitigate the IRS-scattered interference. For example, the BS may apply the ZF precoding, where the beamforming vector for each user $k$, i.e., ${\mv w}_{B,k}$, lies in the null-space of its effective channels with all other $K-1$ users. The optimal multi-beam routing solution can be obtained by conducting a full enumeration over all feasible solutions, while the recursive partial enumeration algorithm in \cite{mei2021mbmh,mei2021massive} can still be applied to balance the performance and complexity. It is also worth noting that (P2) may become infeasible as the number of users is large or some users are close to each other in location. To resolve this issue, each IRS may switch its served user over different time slots, which thus requires a proper user scheduling policy. Alternatively, more sophisticated beamforming and beam routing with relaxed path separation constraints can be designed. As will be discussed later in Section \ref{training}, if practical discrete beamforming codebooks are employed at the BS and each IRS, this may be achieved based on proper beam training design.  

\begin{figure*}[!t]
\centering
\subfigure[3D plot.]{\includegraphics[width=0.35\textwidth]{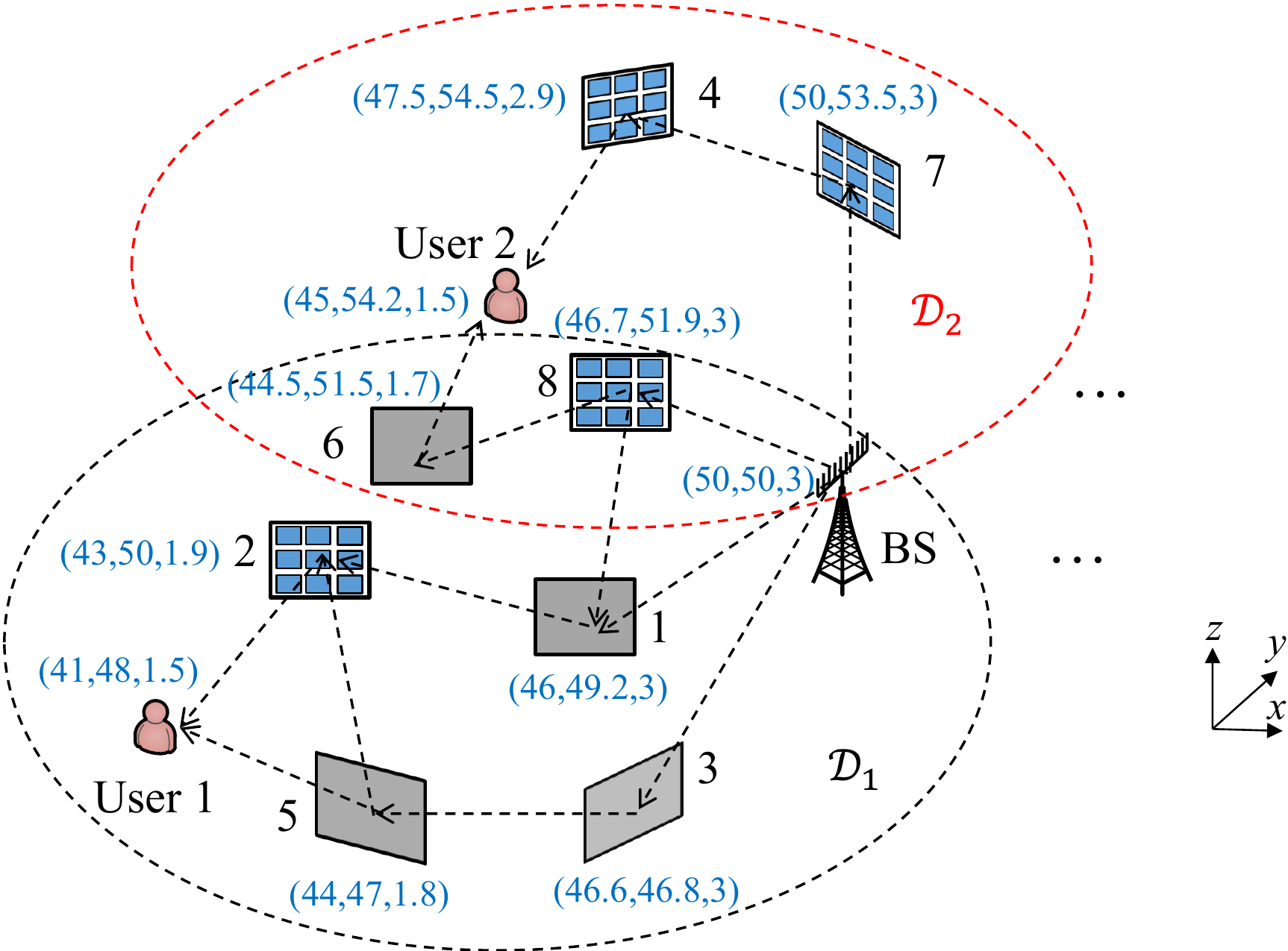}}
\subfigure[LoS graph of the system.]{\includegraphics[width=0.35\textwidth]{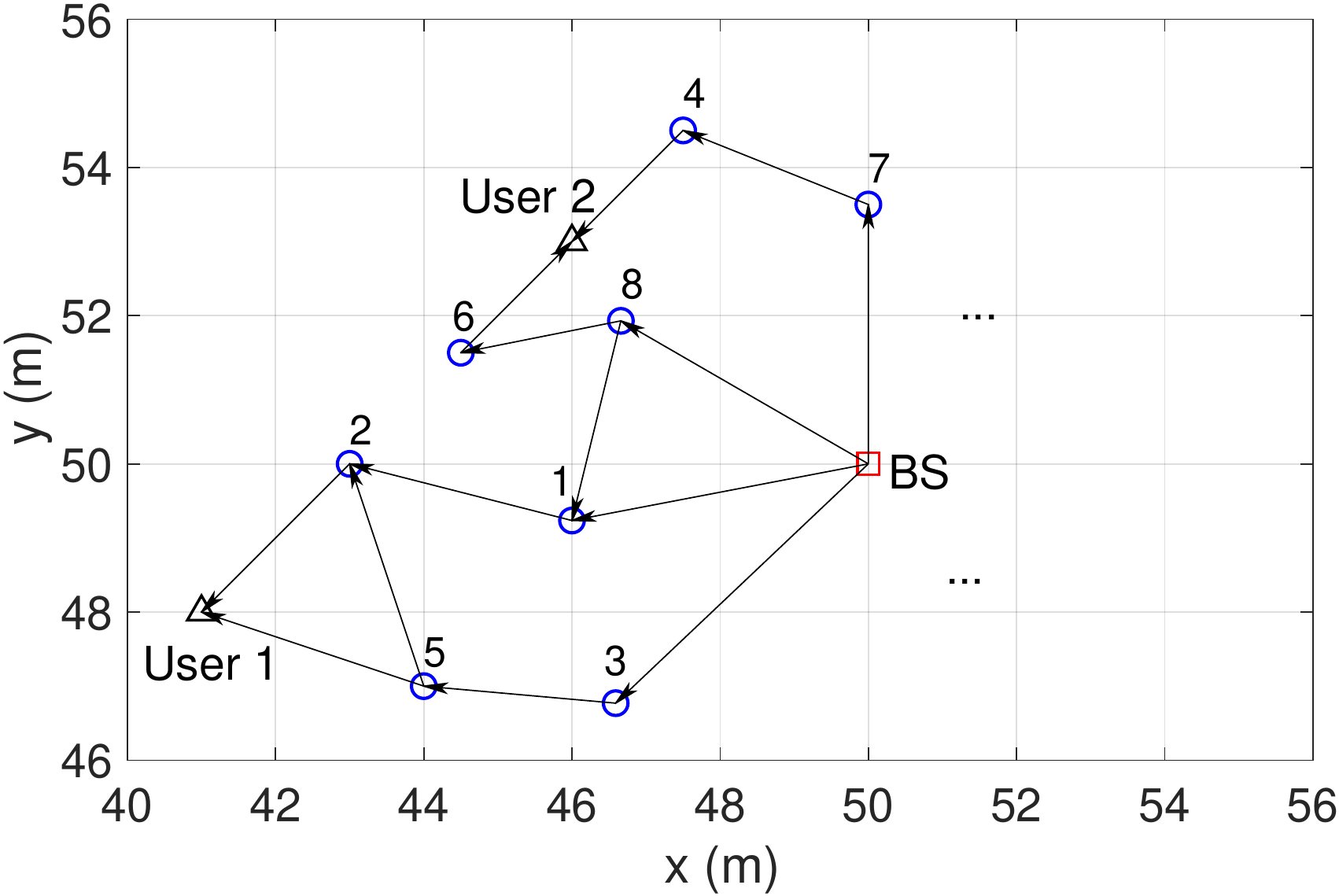}}
\caption{Simulation setup for the multi-IRS multi-reflection system.}\label{topology}
\vspace{-6pt}
\end{figure*}
\begin{figure*}[!t]
\centering
\subfigure[$M_0=20$.]{\includegraphics[width=0.35\textwidth]{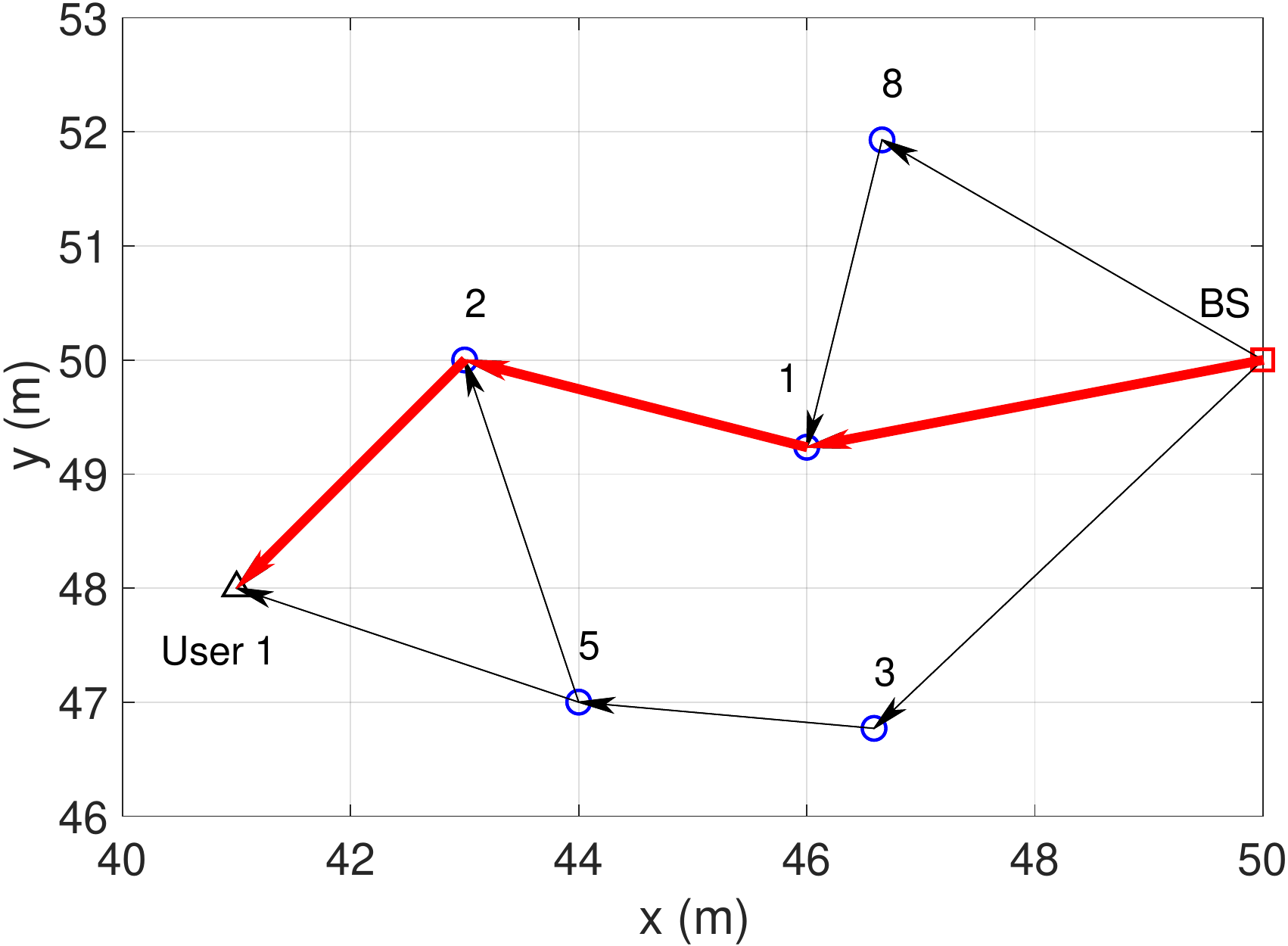}}
\subfigure[$M_0=24$.]{\includegraphics[width=0.35\textwidth]{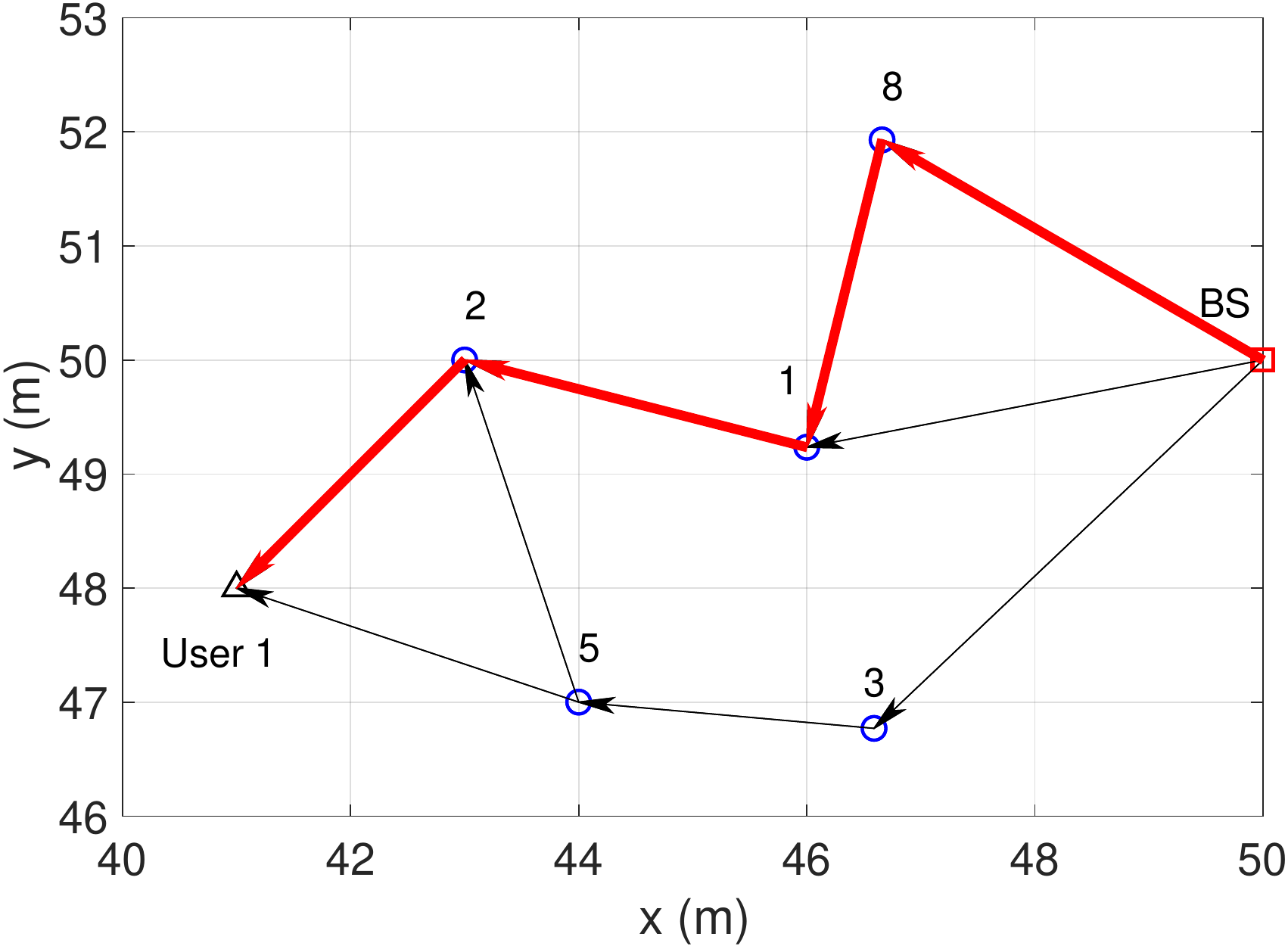}}
\caption{Optimal reflection paths for user 1 only.}\label{SigRefPath}
\vspace{-6pt}
\end{figure*}
\begin{figure*}[!t]
\centering
\subfigure[Optimal reflection paths without the constraint (\ref{op3b}).]{\includegraphics[width=0.35\textwidth]{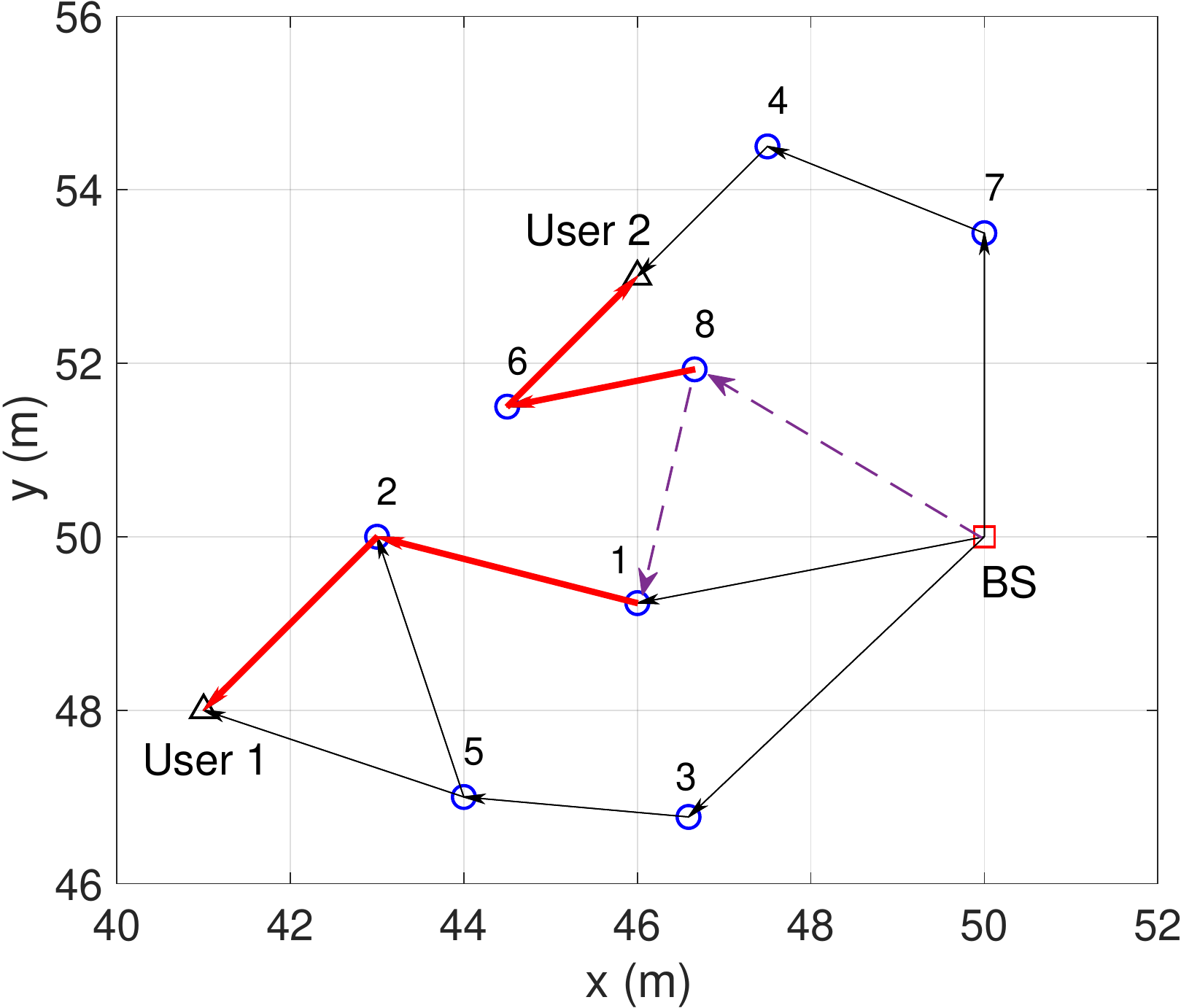}}
\subfigure[Optimal reflection paths with the constraint (\ref{op3b}).]{\includegraphics[width=0.35\textwidth]{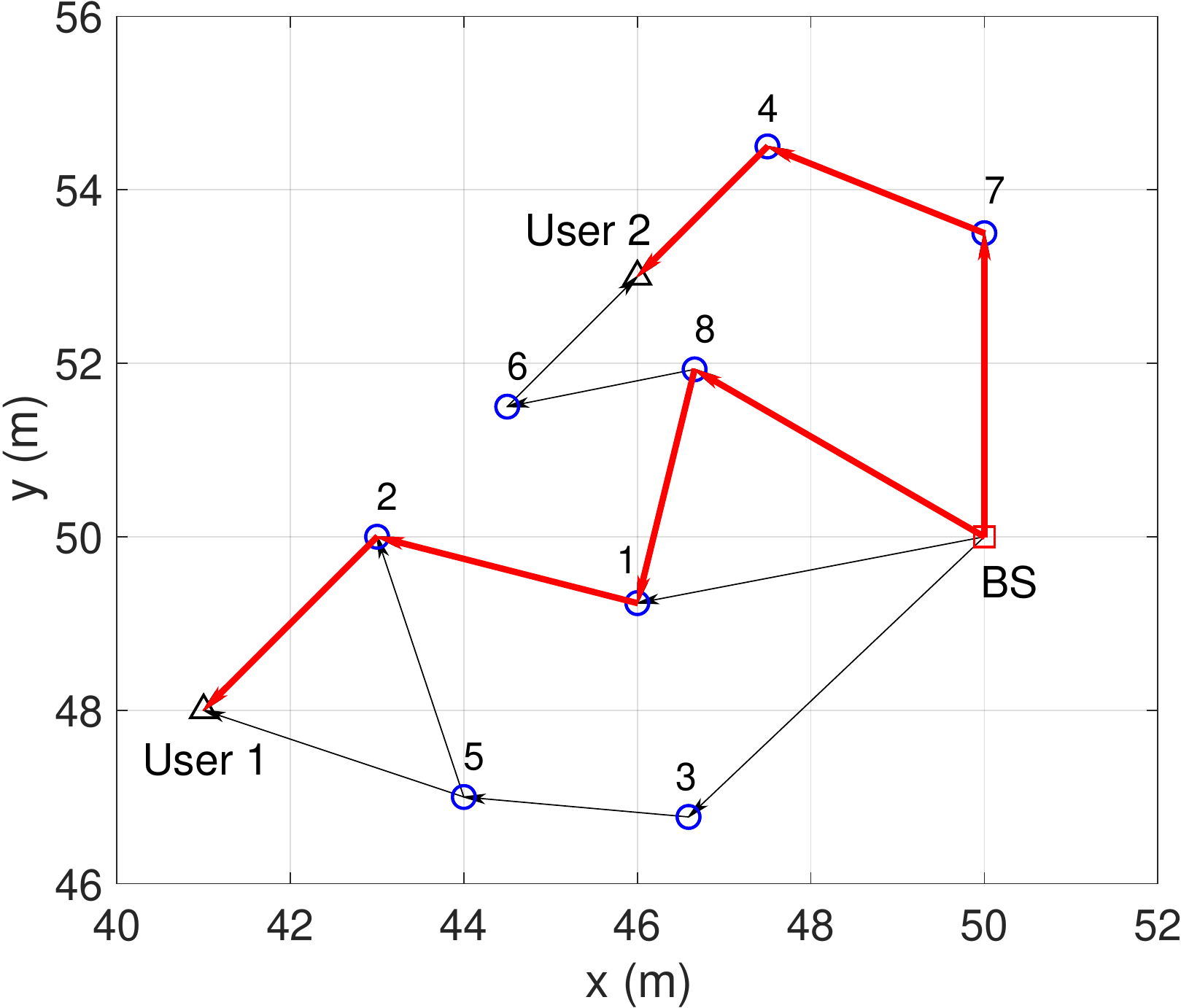}}
\caption{Optimal reflection paths in the presence of both users 1 and 2.}\label{MultRefPath}
\vspace{-12pt}
\end{figure*}
To show the efficacy of the proposed beam routing designs and validate our theoretical analysis, we provide simulation results pertaining to an indoor multi-IRS aided system with two users whose direct channels with the BS are blocked, as shown in Fig.\,\ref{topology}(a). Given the two users' effective IRS regions ${\cal D}_1$ and ${\cal D}_2$, we only need to consider $J=8$ IRSs shown in Fig.\,\ref{topology}(a). The LoS graph of this system is shown in Fig.\,\ref{topology}(b). Each LoS link in the system is assumed to follow Rician fading with a Rician factor of $\kappa$. The number of BS antennas is $N_B=32$. The carrier frequency is set to $f_c=5$ GHz. The antenna/element spacing at the BS/IRS is assumed to be half/quarter-wavelength. The numbers of elements in each IRS's horizontal and vertical dimensions are assumed to be equal and denoted as $M_0$, i.e., $M=M_0^2$. 

First, by assuming that only user 1 exists in the system, Figs.\,\ref{SigRefPath}(a) and \ref{SigRefPath}(b) show its optimal reflection paths under $M_0=20$ and $M_0=24$, respectively, with $\kappa=20$ dB. It is observed that with $M_0=24$, the optimal reflection path goes through one more IRS as compared to that with $M_0=20$. This is due to the different dominating effects of the CPB gain and the end-to-end path loss, as shown in (\ref{eq1}). In particular, when $M_0=20$, the latter dominates the former and thus, the optimal reflection path goes through the least number of IRS reflections. However, when $M_0$ increases to 24, the former is enhanced and dominates the latter. As a result, the optimal reflection path goes through one more IRS to achieve a higher CPB gain. 
Next, in the presence of both users 1 and 2, we plot in Fig.\,\ref{MultRefPath}(a) their respective optimal reflection paths if there is no path separation constraints in (\ref{op3b}) under $M_0=24$. It is observed that the two reflection paths have a common constituent link (from the BS to IRS 8) and there exists an LoS link between them (from IRS 8 to IRS 1), as indicated by the dashed lines, which can result in strong inter-user interference. In Fig.\,\ref{MultRefPath}(b), we plot the optimal reflection paths for the two users under the path separation constraints (\ref{op3b}). By comparing Fig.\,\ref{MultRefPath}(b) with Fig.\,\ref{MultRefPath}(a), it is observed that the reflection path for user 2 is changed; as a result, its rate performance is sacrificed in order to balance the two users' effective channel gains with the BS.\vspace{-8pt}

\subsection{Beam Training}\label{training}
In the previous two subsections, we have shown how to optimize the BS and IRS beamforming in a general multi-IRS aided system if perfect CSI on all LoS links is available, which, however, is practically difficult, if not impossible, to obtain in multi-IRS systems. To facilitate the practical implementation, beam training turns out to be a viable solution, where the BS and each IRS employ a predefined active/passive beamforming codebook consisting of a finite number of beam patterns or directions. As such, the optimal BS and IRS beamforming design can always be obtained via an exhaustive beam search, without the need of explicit CSI or channel estimation.

Let ${\cal W}_B$ and ${\cal W}_I$ denote the active and passive beamforming codebooks at the BS and IRS, respectively. Then, we have ${\mv w}_{B,k} \in {\cal W}_B, k \in \cal K$ and ${\mv \theta}_j \in {\cal W}_I, \forall j \in \cal J$. First, the BS can perform active beam training involving all users, as in the conventional multi-user system. The IRSs in this phase can be simply switched off. By this means, the BS can identify the users that already achieve strong direct channels with it, and these users can be scheduled in the time slots different from other users. Whereas for the users whose direct channels with the BS are insufficiently strong, the BS should further exploit the passive beamforming of IRSs to improve their performance, based on different beam training approaches as detailed next.

{\it 1) Exhaustive and Sequential Beam Search:} Based on the LoS channel model in (\ref{LoSCh}), the optimal active and passive beamforming (${\mv w}_{B,k}, k \in \cal K$ and ${\mv \theta}_j, j \in \cal J$) that maximizes the minimum SINR among all users can be obtained according to the following beam training procedure. In particular, the BS coordinates with all IRS controllers in $\cal J$ to traverse all possible combinations of the active and passive beam patterns by training. In the meanwhile, each user measures its receive SINR for each combination and reports all of its measured SINRs to the BS. Then, the BS can determine the optimal combination of beam patterns by computing and comparing the minimum user SINRs for all beam combinations. However, the total number of combinations of beam patterns for training is $D_BD_I^J$. As such, the optimal beam training via exhaustive beam search will incur prohibitively high complexity and delay for real-time implementation, especially when the number of IRSs $J$ and/or beam patterns $D_B$/$D_I$ is practically large. Alternatively, to reduce its high complexity, the beam patterns at the BS and all IRSs can be sequentially updated until the convergence is achieved. However, this also results in a high complexity in the order of $D_B+JD_I$. Note that in the above two beam training methods, no beam routing is considered explicitly as all IRSs simply search their beam patterns without being assigned for any user exclusively.

On the other hand, if only a single LoS reflection path is selected for each user, then the beam training can be conducted jointly with optimizing the beam routing for each user. In particular, for a given set of reflection paths of the $K$ users (not necessarily separated), exhaustive or sequential beam search can be performed to determine the active/passive beam patterns at the BS/involved IRSs for this path set. Then, similar procedures are conducted for other sets of reflection paths, until all path sets are searched. It is evident from the above that the exhaustive and sequential beam search can both incur exorbitant overhead due to real-time training, with or without considering the beam routing. To tackle this issue, we further propose a new distributed beam training method for the case of multi-beam routing as follows. 

{\it 2) Distributed Beam Training:} The distributed beam training method is composed of offline and online phases, by exploiting the approximately time-invariant BS-IRS and inter-IRS channels and the coordination among the IRS controllers and BS. In particular, the BS and each IRS maintain a local beam training table (BTT) by coordinating with their respective neighboring nodes, and all IRSs' local BTTs are fed back to the BS for determining the joint beamforming and beam routing solution, as elaborated next. 

\begin{figure}[!t]
\centering
\includegraphics[width=3.6in]{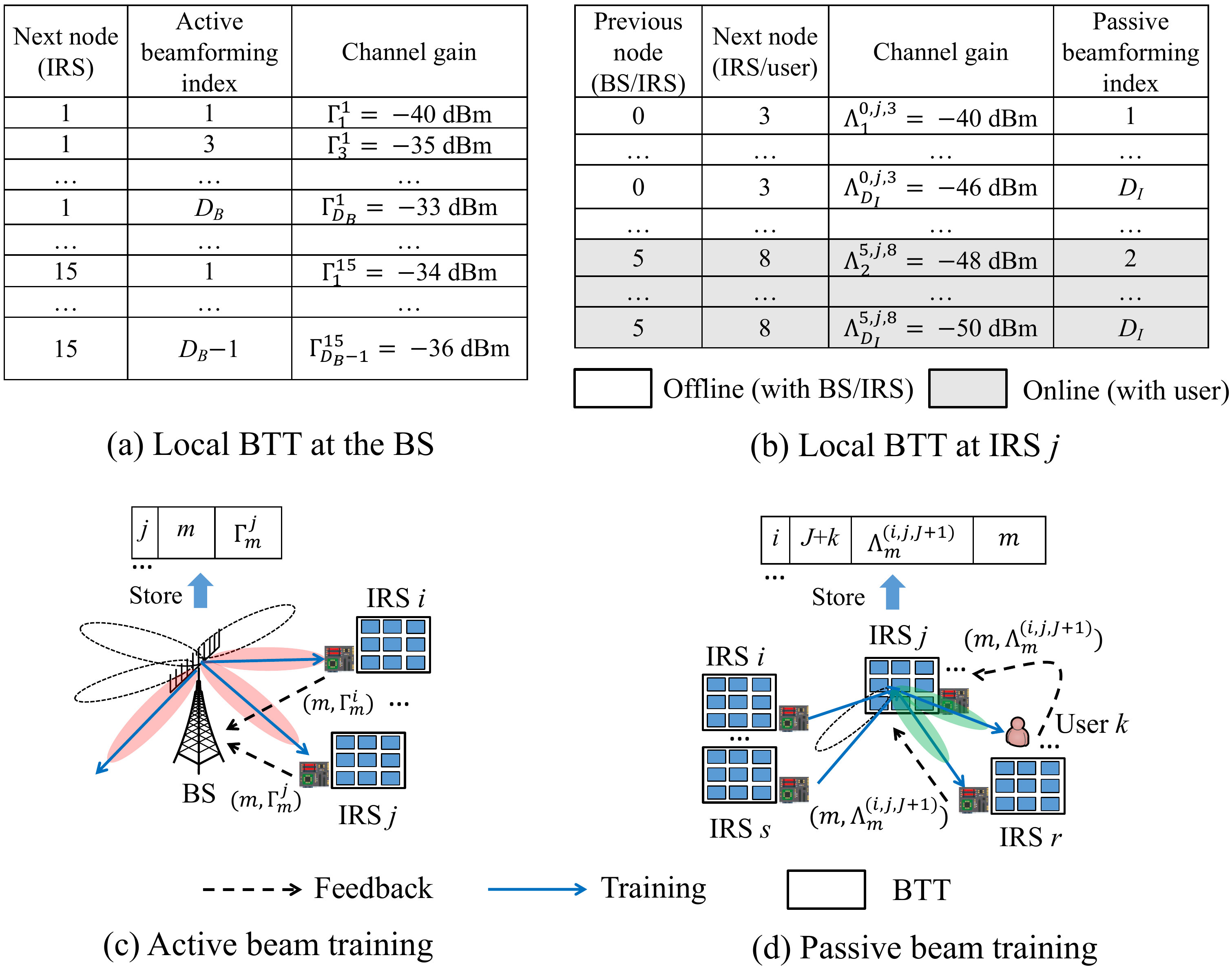}
\DeclareGraphicsExtensions.
\vspace{-9pt}
\caption{Illustrations of BTTs at the BS and IRS and the proposed distributed beam training\cite{mei2021distributed}.}\label{train_BS}
\vspace{-12pt}
\end{figure}

Motivated by the results in (\ref{pbv}) and (\ref{abv}), in the case of LoS channels, the passive beamforming of each IRS is only related to its previous and next reflecting nodes, while the active beamforming of the BS only depends on its next nodes. As such, the BTT at each IRS $j$ specifies the received signal strength (RSS) at each of its next nodes (IRS controller/user) when it reflects the signal from each of its previous nodes (BS/IRS controller) with different passive beamforming vectors in ${\cal W}_I$. Similarly, the BTT at the BS specifies the RSS at each of its next nodes (IRS/user) when it transmits with different active beamforming vectors in ${\cal W}_B$. An example of the BTT at the BS and IRS $j$ is shown in Figs.\,\ref{train_BS}(a) and \ref{train_BS}(b), respectively. Let ${\cal N}^{(P)}_j$ and ${\cal N}^{(N)}_j$ denote the sets of previous and next nodes of IRS $j$ in the beam routing, respectively. The set of next nodes of the BS is denoted as ${\cal N}_0$. Note that the above sets can be determined during the offline/online coordination among the nodes. 

To construct the BTT at the BS, as shown in Fig.\,\ref{train_BS}(c), the BS consecutively sends training symbols from each direction defined in ${\cal W}_B$. In the meanwhile, each IRS controller $j, j \in {\cal N}_0$ measures its average RSS over time for each active beam direction. If its RSS for an active beam is high, this indicates that IRS $j$ may receive strong desired signal (or interference) from this beam in data transmission. Thus, if the RSS is larger than a prescribed threshold, IRS controller $j$ should feed its value back to the BS, which is then stored in the BTT at the BS, as shown in Fig.\,\ref{train_BS}(a). As the BS-IRS channels can be assumed to be constant over a long period, the active beam training can be performed at the BS offline in a periodic manner. 
While to construct the BTT at each IRS $j, j \in \cal J$, as shown in Fig.\,\ref{train_BS}(d), IRS controller $j$ informs each node $i, i \in {\cal N}^{(P)}_j$ (an IRS controller or the BS) to send training signal over consecutive time slots, while it changes the reflection direction based on ${\cal W}_I$ sequentially over time slots. Meanwhile, each node $r, r \in {\cal N}^{(N)}_j$ (an IRS controller or a user) measures the average RSS for different reflection directions in ${\cal W}_I$. Similarly as the active beam training, if the RSS is larger than a prescribed threshold, node $r$ should report it to IRS controller $j$, which then stores it in the BTT, as shown in Fig.\,\ref{train_BS}(b). Note that the passive beam training without involving any user, i.e., $r<J+1$, can be conducted offline, thus greatly reducing the online training overhead as compared to the exhaustive/sequential beam training previously described.

Each IRS controller $j, j \in \cal J$ feeds back its BTT obtained offline to the BS. If one or more users are discovered in its neighborhood, it also sends back the corresponding new entries in its BTT (see, e.g., the last several rows of Fig.\,\ref{train_BS}(b)) obtained online to the BS. Based on all the BTTs from the IRS controllers and its own BTT, the BS can construct a global BTT and estimate an approximated value of its effective channel gain with each user $k$ under any reflection path $\Omega^{(k)}$, as well as the active and passive beamforming applied by itself and involved IRSs, respectively. In particular, the approximated value would become more accurate if the LoS propagation is dominant in each constituent link of the reflection path $\Omega^{(k)}$, as shown in Section \ref{MUMR}. Based on the approximated channel gains, the BS can properly select the beam patterns and reflection path for each user to ensure the strength of its desired signal while controlling the (worst-case) interference power from all other paths with it. Nonetheless, as the distributed beam training only accounts for partial CSI involving the IRS controllers in the system, it may only yield a suboptimal beamforming and beam routing solution, while significantly reducing the complexity and overhead of real-time beam training. Finally, the BS sets its active beamforming as the optimized one and sends the indices of its optimized passive beam patterns to different IRS controllers via the downlink control links, which then tune the passive beamforming of their respectively controlled reflecting elements accordingly.

\begin{figure}[!t]
\centering
\includegraphics[width=3.2in]{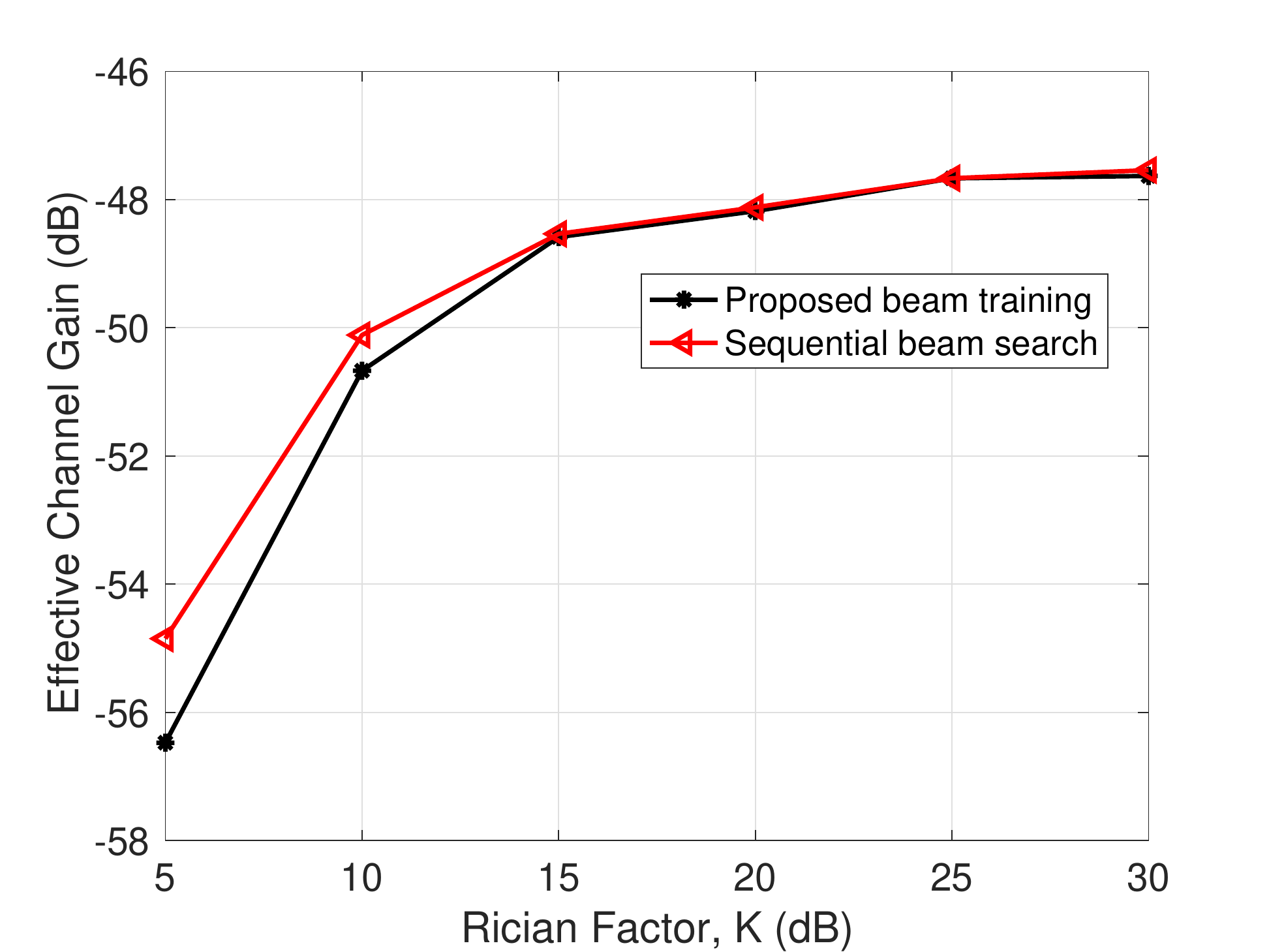}
\DeclareGraphicsExtensions.
\caption{Effective channel gains by two beam training schemes versus Rician factor $\kappa$, $M_0=24$.}\label{rician}
\vspace{-12pt}
\end{figure}
To evaluate the efficacy of the distributed beam training, we next present a case study assuming the single-user system. In this case, based on the global BTT at the BS, it can compute the beamforming design that maximizes the approximated channel gain under any given BS-user reflection path, thereby computing the optimal reflection path or beam routing solution by solving an optimization problem similar to (P1)\cite{mei2021distributed}. For example, consider the reflection path shown in Fig.\,\ref{SigRefPath}(b). Under different Rician factors $\kappa$ and $M_0=24$, we show its effective channel gains by the distributed beam training and the sequential beam search, respectively, in Fig.\,\ref{rician}. Here, we assume that each IRS employs three-dimensional (3D) passive beamforming and use 32-point discrete Fourier transform (DFT)-based codebooks \cite{mei2021massive} at the BS and each IRS's two dimensions (horizontal and vertical). It is observed from Fig.\,\ref{rician} that the channel gains by both schemes increase with $\kappa$. This is expected since the LoS component between any two consecutive nodes in the reflection path becomes more dominant as $\kappa$ increases, which enhances both the active beamforming gain and CPB gain. It is also observed that their performance gap decreases with $\kappa$ and becomes negligible as $\kappa \ge 15$ dB. This is due to the fact that the approximation accuracy of the distributed beam training improves with increasing $\kappa$. \vspace{-6pt}

\subsection{Other Related Work and Future Direction}
The research on the general multi-IRS system with multi-reflection links is still in its infancy. As such, we next point out several interesting and promising directions worth further investigating in future work.

{\it 1) IRS Deployment/Placement:} Similar to the double-IRS system, IRS deployment also has a great impact on the performance of multi-IRS systems. In particular, IRSs should be densely deployed in the region of interest to shorten their mutual distances as well as the distances from the BS/users, so as to maximize the number of LoS links and ensure sufficient path diversity. In addition to the inter-node distances, the facing direction of each IRS also affects the performance, as it determines the number of “effective” LoS links due to the half-space reflection limitation of each IRS. Furthermore, in the case of discrete beamforming codebooks, if the resolution of the codebooks is low or moderate, the maximum CPB gain in (\ref{eq1}) cannot be fully achieved and instead, it critically depends on the angles of arrival and departure between nearby IRSs\cite{mei2021mbmh}. Despite of the above issues, fortunately, the IRS deployment can be designed offline based on historical performance data collected in the network.

{\it 2) Broadband System with Frequency-Selective Channels:} In our previous discussion, we consider the narrowband system with frequency-flat channels for simplicity. In the more general broadband system, a common passive beamforming should be designed at each IRS to cater to multiple frequency bands of each user with different channels in general, due to the lack of baseband processing capability at the IRS. This makes the beamforming and beam routing design problem more challenging to solve. A practical approach to solve this problem is that each IRS splits its elements into multiple sub-surfaces (or equivalently, co-located smaller IRSs), each optimizing its reflection for the signal in one frequency band only. The path separation constraints introduced in Section \ref{MUMR}2 can be similarly applied to those co-located IRSs reflecting user signals over the same frequency band to mitigate the co-channel inter-user interference, while the other co-located IRSs can be treated as environment scatterers. In addition, the proposed distributed beam training scheme can also be applied in the case of frequency-selective channels. However, each IRS controller may need to maintain more BTTs, each for one co-located IRS/frequency band. As a result, the IRS beam routing/passive reflection design becomes more challenging to solve, which is  worthy of further investigation. 

{\it 3) Hybrid Active/Passive IRSs for Multi-Reflection:} As shown in (\ref{eq1}), for a multi-reflection link, the number of elements at each IRS, $M$, needs to be sufficiently large to compensate for the multiplicative path-loss. However, this inevitably incurs higher hardware and energy cost for each IRS. Recently, a new type of IRS, termed active IRS, has been proposed to overcome the above issue of passive IRS by reflecting and amplifying the incident signal at the same time with low-cost hardware\cite{long2021active,zhang2021active,you2021active}. Different from the conventional amplify-and-forward relay, it directly reflects signals in a full-duplex (FD) manner, thus achieving high spectral efficiency. It was shown in \cite{you2021active} that an active IRS is able to yield a stronger strength of the reflected link than a passive IRS if its amplification gain is sufficiently large and/or the number of its elements is small-to-moderate. As such, it may be more suitable to multiple signal reflections than the passive IRS if the number of IRS elements is not large. It is thus interesting to study the beamforming and beam routing problem for active IRSs and characterize its performance gain over our considered system with passive IRSs only. It is also interesting to combine active and passive IRSs by leveraging their complementary benefits for optimizing the multi-reflection link performance.

{\it 4) Cascaded Channel Estimation for Multi-Reflection Links:} In Section \ref{training}, we have discussed the distributed beam training to achieve low-complexity beamforming design without the need of explicit CSI or channel estimation, which, however, generally comes at the expense of beamforming performance degradation, especially if the LoS propagation is not sufficiently dominant (e.g., when the Rician factor $\kappa=5$ dB in Fig.\,\ref{rician}). However, to enable the optimal beamforming design for the multi-IRS system, accurate CSI is required but more challenging to acquire in practice as compared to single-/double-IRS systems.
This is because, in the case of fully passive IRSs, it is practically infeasible to acquire the separate CSI between any two IRSs as well as that between IRS and BS/users. Moreover, it is not straightforward to directly apply the existing cascaded channel estimation methods for single-/double-reflection links to the multi-reflection link, since the number of cascaded channel coefficients grows exponentially with the number of IRS reflections, as discussed in Section~\ref{SU_MISO_CE}.
One possible strategy to overcome this difficulty can be decomposing the high-dimensional cascaded channel into a series of lower-dimensional sub-channels that are easier to estimate with lower training overhead. 
Towards this end, new channel estimation algorithms and training designs (including both pilot sequence and multi-IRS training reflection pattern) by exploiting some special properties (such as common inter-IRS channels) of the high-dimensional multi-reflection channel are highly desired, which is worthy of further investigation in the future.

\section{Conclusions}
In this paper, we provide a tutorial overview of multi-IRS aided wireless networks with two or more IRSs employed for assisting each wireless link, which is shown to provide promising performance gains over the conventional single-IRS aided system under certain conditions. To this end, we first present the general system model for a multi-IRS aided wireless network. Then we review the state-of-the-art results on the double- and multi-IRS aided systems, respectively, by focusing on addressing their new and unique challenges in the design of  IRS reflection optimization and channel acquisition, as well as highlighting open issues and important directions for future research. Besides IRS-aided communications, there are also other interesting and promising wireless  applications that can benefit from the multi-IRS aided wireless network, such as wireless power transfer, RF sensing/localization, spatial modulation, etc., which require further investigation in future work as well. For example, the promising CPB gain of multi-IRS reflection  can be exploited to enhance the range and  efficiency of RF-based wireless power transfer. It is hoped that this paper will serve as a useful and inspiring resource for future research on IRS  to unlock its full potential in future-generation (B5G/6G) wireless communications.

\bibliography{Multi-hop_tut}
\bibliographystyle{IEEEtran}

\end{document}